\documentclass[pre,reprint,groupedaddress,superscriptaddress]{revtex4-1}
\usepackage{amsmath,amsthm}
\usepackage{graphicx}

\let\textcite\relax
\makeatletter
\DeclareRobustCommand{\MakeUppercase}[1]{{%
      \def\i{I}\def\j{J}%
      \def\reserved@a##1##2{\let##1##2\reserved@a}%
      \expandafter\reserved@a\@uclclist\reserved@b{\reserved@b\@gobble}%
      \protected@edef\reserved@a{\uppercase{#1}}%
      \reserved@a
  }}
\DeclareRobustCommand{\MakeLowercase}[1]{{%
      \def\reserved@a##1##2{\let##2##1\reserved@a}%
      \expandafter\reserved@a\@uclclist\reserved@b{\reserved@b\@gobble}%
      \protected@edef\reserved@a{\lowercase{#1}}%
      \reserved@a
  }}
\makeatother
\expandafter\let\csname ver@natbib.sty\endcsname\relax

\usepackage[style=numeric-comp,sorting=none,backend=bibtex]{biblatex}

\usepackage{hyphenat}

\usepackage{nameref}
\newcommand{\beginsupplement}{%
\setcounter{table}{0}    \renewcommand{\thetable}{S\arabic{table}}     \setcounter{figure}{0} \renewcommand{\thefigure}{S\arabic{figure}}
}

\begin{document}
\renewcommand{\bibliography}[1]{}

\author{Jeff Alstott*}
\email{alstott@mit.edu}
\affiliation{Massachusetts Institute of Technology}
\affiliation{Singapore University of Technology and Design}
\author{Giorgio Triulzi*}
\email{gtriulzi@mit.edu}
\affiliation{Massachusetts Institute of Technology}
\affiliation{Singapore University of Technology and Design}
\affiliation{United Nations University - MERIT}
\author{Bowen Yan}
\affiliation{Singapore University of Technology and Design}
\author{Jianxi Luo}
\affiliation{Singapore University of Technology and Design}

\title{Mapping Technology Space by
Normalizing Patent Networks}

\begin{abstract}
Technology is a complex system, with technologies relating to each other in a space that can be mapped as a network. The technology network's structure can reveal properties of technologies and of human behavior, if it can be mapped accurately. Technology networks have been made from patent data, using several measures of proximity. These measures, however, are influenced by factors of the patenting system that do not reflect technologies or their proximity. We introduce a method to precisely normalize out multiple impinging factors in patent data and extract the true signal of technological proximity, by comparing the empirical proximity measures with what they would be in random situations that remove the impinging factors. With this method, we created technology networks, using data from 3.9 million patents. After normalization, different measures of proximity became more correlated with each other, approaching a single dimension of technological proximity. The normalized technology networks were sparse, with few pairs of technology domains being significantly related. The normalized network corresponded with human behavior: we analyzed the patenting histories of 2.8 million inventors and found they were more likely to invent in two different technology domains if the pair was closely related in the technology network. We also analyzed 250 thousand firms' patents and found that, in contrast, firms' inventive activities were only modestly associated with the technology network; firms' portfolios combined pairs of technology domains about twice as often as inventors. These results suggest that controlling for impinging factors provides meaningful measures of technological proximity for patent-based mapping of the technology space, and that this map can be used to aid in technology innovation planning and management.

\end{abstract}

\maketitle

\section{Introduction}
Technological invention can be considered as navigating a space of technologies \cite{Kauffman2000, Strumsky2002, Fleming2004, Silverberg2005, Frenken2006, Silverberg2007}. Networks have been used to represent and describe that space: there are many kinds of technologies, and they relate to each other in many, complex ways \cite{Kay2014, Leydesdorff2014, Breschi2003}. Different technological domains can be connected and proximate in the technology space if they rely on similar or related knowledge \cite{Leydesdorff2014, Arthur2009, Shiffrin2004,Mane2004, Verspagen1997a, Jaffe1986}. An accurate network map of the technology space, even at a low resolution, opens the door to understanding how technology as a whole behaves and how humans interact with it. Such understanding could improve the inventive strategies of individual inventors and the technology innovation policies of firms or countries.

The technology space can be mapped using patent data, and technology domains identified from patent metadata. Domain experts at patent offices classify every patent into one of many technology classes, such as ``organic chemistry" or ``hats," which represent technology domains. The proximity between two technology classes can be measured in numerous ways using patent data, and different measures reflect different intuitions of how technologies could be related or similar with each other. This creates two types of challenges when mapping the space of technologies. First, when different measures disagree on the proximity of two technology domains, it can make interpretation of the technology space difficult. This makes it harder to derive precise technology development strategies based on the network map of the technology space. If different methods for quantifying proximity could be harmonized, it would allow for greater clarity in studying technology. Second, the different empirical measures of inter-domain distance in the technology space are affected by biases that arise from the patenting and inventive processes, which distort the perceived proximity across technologies. Hence, these measures needs to be properly cleaned to allow the true representation of the technology space to emerge.

\begin{figure*}[ht!]
\begin{center}
\includegraphics[width=\textwidth]{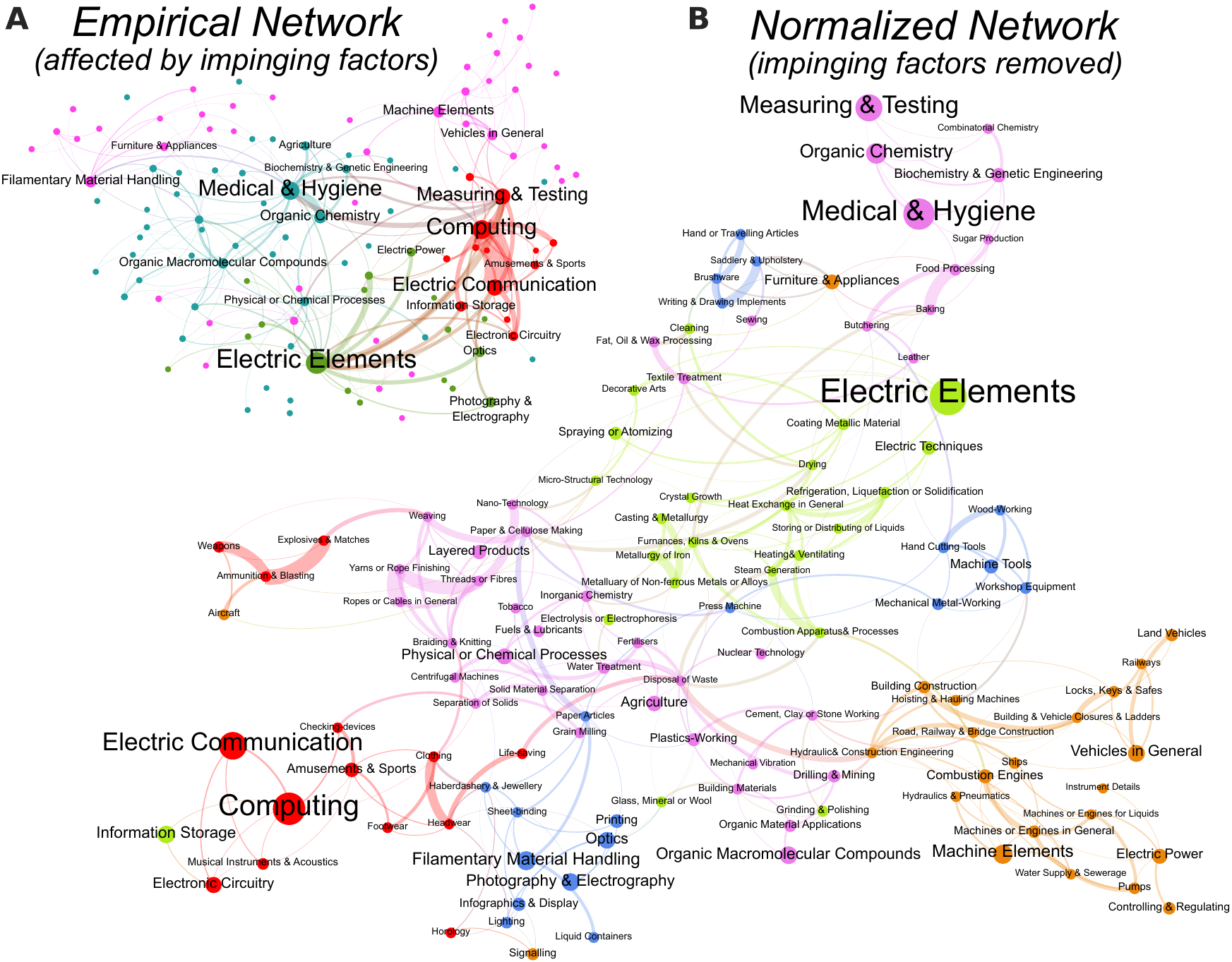} 
\end{center}
\caption{\textbf{Measuring technology networks that control for impinging factors reveals different network structures.} A) The empirical and B) normalized technology networks. The normalized network's links are not influenced by the impinging factors of Fig. \ref{Impinging_Factors}. Node size: proportional to the number of patents in the technology class. Link weight: proportional to the number of citations between the technology classes (the average of both directions). The networks are fully connected, but only a subset of the strongest links are visualized (see \textit{Appendix} \ref{network_visualization}). For visual reference, a community structure is shown by node color, which was identified using both visualized and unvisualized links.
}\label{Empirical_Normalized_Networks}
\end{figure*}

We used data from 3,911,050 utility patents issued from 1976 to 2010 by the United States Patent and Trademark Office to create technology networks using nine different measures of proximity. Fig. \ref{Empirical_Normalized_Networks}A shows one such network. Each of the proximity measures builds on one of two phenomena: 1) patents' citations to other patents in different technology classes, or 2) classes occurring together on patents or in the patenting histories of inventors or firms. Both of these phenomena, are influenced by impinging factors that are not intrinsic properties of the technologies that the patents represent (Fig. \ref{Impinging_Factors}). These factors include classes' number of patents, the number of citations those patents make, and how old the patents are; we examine their origins below. These factors distort the perception of how distant two technologies are in the space based on raw empirical measurements. The accuracy of a technology network map crucially depends on the ability to separate the true signal of technological proximity from these other, spurious effects. By removing the spurious effects, we can create a better representation of the latent technology network (Fig. \ref{Empirical_Normalized_Networks}B).

Here we introduce a method to precisely control for the complexities of multiple impinging factors in patent data, all at once. We calculated a null hypothesis: an expectation of what the observed proximity measure between two technology classes would be by random chance, given the other influencing factors. We then identified which pairs of technology classes had significantly higher proximity than that expected by chance. After normalizing the empirical proximity measures relative to the random expectation, most measures of technology proximity became highly correlated with each other. Normalized proximity measures based on citations, co-classification, and inventors' patent portfolios are all correlated strongly. The sole measure of technology class proximity with lower correlation to the others was how often firms patented in a pair of classes, which could be driven by factors beyond how proximate the technologies are. These results indicate that controlling for the impinging factors creates stronger agreement of different measures of proximity, validating the utility of the method. The increased agreement of measures also opens up the possibility of measuring a single technology space and constructing a unified patent technology map.

\section{Measuring Technology Proximity}\label{measuresreview}
The proximity of technologies has been measured in various ways. Here we briefly review the most common proximity measures. The proximity measures can be organized into two families based on the kind of data they use: citations and co-occurrences.

\subsection{Citation based measures}
Patents cite other patents as related technologies. Citations highlight which existing solutions the current invention has improved upon, with the purpose of limiting what the citing patent can claim as novel intellectual property. Citations can thus represent knowledge proximity. Several measures of knowledge proximity between technology classes have been created in the literture, building upon patent citations.

\subsubsection{Direct Citation}
The most straightforward way to describe the proximity between two technology classes is to simply count the number of citations between them \cite{Leten2007}. The Direct Citation measure is the total number of citations from patents in a class $X$ to other patents in another class $Y$. Because citations disclose the relevant prior art, the direct citation count between classes can be interpreted as an overall measure of the importance of the cited class as a technical input for the citing one. 

\subsubsection{Co-Citation}
Patents can make many citations, including to patents from multiple classes. If two classes are often cited together they may function well together as a input. The Co-Citation between two classes $X$ and $Y$ is the number of patents that cited patents from both $X$ and $Y$. Co-Citation thus indicates if two classes often jointly serve as knowledge inputs for the same inventive output. Co-Citation has been used to measure the proximity of scientific fields and journals \cite{Uzzi2013, Wallace2009}. Co-Citation is sometimes normalized by computing the Jaccard index (dividing the number of co-citations by the total number of citations received by patents in the two classes $X$ and $Y$ \cite{SmallHenry1973}), but here we calculate Co-Citation directly and use more complex normalizations, as described below.

\subsubsection{Cosine Similarity}
A more sophisticated measure of proximity is not whether two classes cite each other, but if they cite other classes in a similar pattern (i.e. if they use the same set of inventive inputs). This is analogous to measuring the structural equivalence of two nodes in a network \cite{Leicht2006}. We count how many citations patents in a class $X$ make to patents in every other class ($Y$, $Z$, and so on). This is summarized as a class-class citation vector, $c_X$. If the class-class citation vector is the same for two classes, they have the same citation behavior, and are taken to be related or proximate. If they have entirely different vectors, they have entirely different citation behaviors, and are taken to be unrelated. We calculate the similarity of the two class-class citation vectors by taking the cosine of the angle between them, $cos(c_X, c_Y)$. 

Cosine Similarity is a long-used measure for evaluating the similarity of two sets. The cosine index was introduced as a measure of proximity of technology domains in patent data by Jaffe \cite{Jaffe1986, Jaffe1989}. Jaffe measured relatedness between pairs of technological fields (proxied by patent classes) by computing the cosine of the vectors representing the occurrences of fields in firms' patent documents. Breschi and colleagues \cite{Breschi2003} designed a similar version of the index, which measures proximity between class pairs as the cosine of the classes' vectors of co-occurrences in patent documents. Cosine similarity has been used in other studies to create patent-based technology maps \cite{Kay2014, Leydesdorff2014}.

Cosine Similarity can be calculated using two different class-class citation vectors: the vector of citations the class $X$ \textit{makes} to every other class, and the vector of citations the class \textit{receives} from every other class. These can be thought of as measuring the similarity of knowledge inputs to the class vs. the similarity of knowledge outputs of the class. We refer to these two measures as Cosine Similarity, Inputs and Cosine Similarity, Outputs.

The principle of measuring class-class citation vectors can be extended to class-patent citation vectors \cite{Yan2015}. In this case, what is measured is how many citations patents in a class $X$ make to every individual patent, without summarizing them by which classes those patents are in. This creates a much higher dimensional vector (as there are many more patents than classes), but the principle is the same. Obviously, the citation vector measured at the class-patent citation level has a higher granularity than its class-class counterpart. This can be interpreted as a measure of similarity of specific, rather than generic, knowledge inputs or outputs between classes. The class-patent citation vectors between two classes can again be compared using Cosine Similarity. Again there are two versions of the measure, depending on whether we measure the citations a class makes versus the citations a class receives. We refer to these two measures as Cosine Similarity, Input , High Resolution and Cosine Similarity, Output, High Resolution. 

\subsection{Co-Classification and Co-Occurrence based measures}
Patents are assigned a main class to which they primarily belong, which is the class used for the citation analysis. But patents are also frequently assigned to additional, secondary classes. We can then measure how often two classes both appear on the same patents together (Co-Classification). This is a common method in scientometric analysis \cite{Jeong2015, Dolfsma2011, Joo2010, Joo2009, Leydesdorff2008, Engelsman1994}. Similarly, an inventor or firm can have multiple patents, and those patents could be in multiple classes. We can then measure how often two classes both appear together in inventors' and firms' patent histories (Co-Occurrence).  Co-Classification is interpreted as measuring how often two technology domains are combined into an invention, while Co-Occurrence is interpreted as how often two technology domains are both used within a single mind or collection of minds (i.e. a firm). Therefore, Co-Classification is a measure of proximity of two technologies based on how similar their artifacts are. In contrast, Co-Occurrence measures the similarity in the technical skills required to make the artifacts, or in the assets or managerial practices needed to be successful in both \cite{Bryce2009, Bottazzi2010, Teece1994}.

Calculating Co-Occurrence from patent data requires accurately tracking individual inventors and firms with multiple patents, even though their names can be listed differently on different patents (e.g. ``IBM'' vs. ``International Business Machines'' or ``Charles Jacob Smith'' vs ``Charles J. Smith''). The recent availability of harmonized inventors' and firms' names for patent data \cite{Li2014} made it possible to compute a reliable measure of Co-Occurrence of technology classes in inventors' and firms' patenting histories. Following the work done in \cite{Yan2015}, this paper represents one of the first attempts at measuring the Co-Occurrence of technology classes in inventors' patenting histories. Inventor and firm identities were tracked across patents using name reconciliation data from \cite{Li2014}. This data identified 2,756,508 inventors and 247,913 firms. Firm identity reconciliation, performed by \cite{Li2014} and based on \cite{Hall2001}, focused on linking patents' assignee names to firms traded in the United States stock market and harmonizing spell variations. The firm identity reconciliation did not merge firms' subsidiaries, which can be distinct entities with different knowledge, capabilities, and operations. Interpreting subsidiaries' relationships with each other is a complex topic that we will not seek to resolve here; we simply consider subsidiaries as separate entities.

\section{Origins of Factors Impinging on the Empirical Measurement of Technology Proximity}

Unfortunately, all the measures of technology proximity are affected by factors other than the technologies themselves. These factors thus impinge on the measures of proximity, detracting from the signal we desire to measure. The impinging factors that we can control for are different depending on if the proximity measure is based on citations or on occurrence data.

\subsection{Citations}
\begin{figure}[ht!]
\begin{center}
\includegraphics[width=\columnwidth]{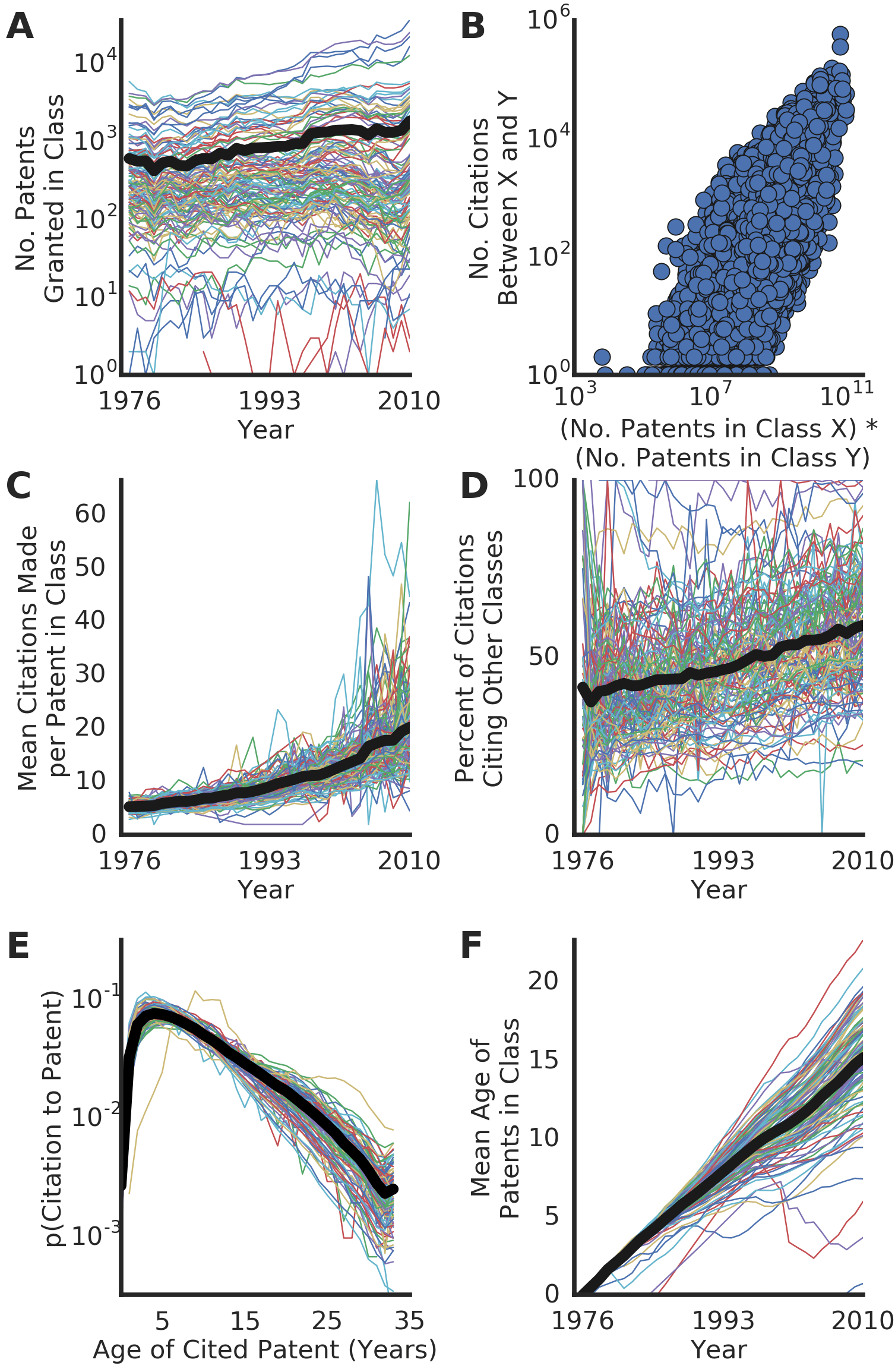} 
\end{center}
\caption{\textbf{Multiple aspects of the patenting system affect citation rates, and thus impinge on measures of proximity.} The impinging factors vary greatly between technology classes and are not stable over time, but these complexities can be removed from measures of technology proximity by normalizing with randomized controls. Colored lines: Individual classes. Black lines: Averages.
A) The number of patents in different technology classes, over time.
B) The number of citations between every pair of classes compared to the number of patents in the two classes (on a double logarithmic scale). On average the two values are proportional.
C) The mean number of citations made by patents in different technology classes, over time. 
D) The percent of citations that cite patents in classes different from that of the citing patent, grouped by the class of the citing patent, over time.
E) The distribution of the ages of patents that are cited, grouped by the year the citation was made.
F) The average age of patents in different technology classes over time. The data for E and F are censored, as they only cover patents awarded from 1976 onward. 
}\label{Impinging_Factors}
\end{figure}

The probability of a citation between two patents, or between two technology classes, is affected by several variables that are not intrinsic properties of the inventions they represent \cite{Hall2001, Uzzi2013}. The expected number of citations between any two technology classes depends on several factors, which vary greatly across classes and time (Fig. \ref{Impinging_Factors}).

First, the expected number of citations between a citing class and a cited class is driven by the number of patents in each (Fig. \ref{Impinging_Factors}A,B). Technology classes vary greatly in size, and those sizes change over time (Fig. \ref{Impinging_Factors}A). All else equal, larger classes both make and receive more citations, and thus there is a linear correlation between the sizes of two technology classes and the number of citations between them (Fig. \ref{Impinging_Factors}B).

Another possible influence of examiner behavior changing over time is the number of citations made per patent. The number of citations made per patent varies across technology classes and history, and is increasing over time (Fig. \ref{Impinging_Factors}C). This may be a cognitive bias due to the growing pool of potential prior art (more previous patents) and patent electronic databases, which makes searching for prior art easier and thus assessing novelty more stringent.  

Patents' citations are also biased to be made to patents within the same class. Citations are often made by patent examiners \cite{Criscuolo2008, Alcacer2006}, and the examiners leverage the classification system to make citations. They first classify the patent, then search for potentially related patents \cite{2014}, and so are more likely to find relevant previous patents to cite within the same class. The portion of citations made to patents in other classes varies greatly depending on the class of the citing patent, and is also growing over time (Fig. \ref{Impinging_Factors}D); these differences can be due to differences in examiner behavior or office policy. 

Another factor affecting citations is time. Recent inventions need time to be recognized and older technologies gradually become unused, though can potentially remain indefinitely \cite{Hall2001,Valverde2007}. Since a patent's citations reflect what technologies are relevant prior art to the invention, these temporal effects are reflected in the citation record: Fig. \ref{Impinging_Factors}E shows the distribution of the age of patents at the time they are cited. Obviously, because the patent data we studied only extends to 1976, the maximum possible patent age depends on the year in which the citing patents was awarded.
The increased likelihood of citing patents from a particular point in history interacts with the growing number of new patents and the increasing number of citations they make. It also crucially interacts with the fact that technology classes vary in age. The average age of patents in a technology class varies across classes, and is increasing over history (Fig. \ref{Impinging_Factors}F). The trend is increasing partly by construction, as we have no data before 1976, but the large variance for recent years is likely a real phenomenon.

Given all these citation phenomena, the expected number of citations between any two technology classes are influenced by their propensity to cite and be cited by other classes, their number of patents, the age distribution of their patents and their propensity to make and receive citations. In this study we show how to simultaneously control for all these factors.

\subsection{Co-Classification and Co-Occurrence}
Like citations, Co-Classification and Co-Occurrence measures are also influenced by other impinging factors, such as the simple number of occurrences. The probability that two technology classes co-occur within the same patent document, inventor's or firm's past patenting history depends on the number of classes that are associated with a patent, inventor or firm and the number of patents, inventors and firms that are associated with a given technology class. A given technology class may be very common or very rare across all patents, inventors, or firms. Similarly, each patent, inventor, or firm may associate with very many technology classes, or very few. 

As explained by Bottazzi and Pirino \cite{Bottazzi2010}, in order to properly measure the true proximity between classes as a function of their Co-Occurrence it is crucial to compare the observed Co-Occurrence with a null hypothesis in which occurrences of classes in patents, inventors' and firms' histories are randomly distributed while preserving \textit{both} the number of occurrences of a class \textit{and} the number of classes that are associated with a given patent, inventor or firm. This is necessary to make sure that the random expectations incorporate the characteristic skewed distributions of the number of classes per patent/inventor/firm and the number of patents/inventors/firms per class that are observed in the real world.

Controlling for the number of occurrences in co-occurrence data has been addressed in information science, ecology, medicine and economics \cite{Eck2009,Ulrich2007,Stone1990,Gobbi2014,Neffke2008}.
We extend on this understanding by also controlling for temporal effects, since the number of occurrences of a class, patent, inventor or firm can vary over time. While some classes are popular in some years and not in others, a change in popularity ought not influence the true proximity between technology classes. For example, if a firm only worked on what was most popular every year, the firm's activity would not provide new information on how technologies are related. Controlling for temporal effects allows us to measure how unusual it is that two classes co-occurred, given \textit{when} they were each popular.

\clearpage
\section{Methods: Measuring Technology proximity \\while Controlling for Impinging Factors}

\subsection{Citations}
Using citation information in patent documents, we measured how often patents in two classes cited each other directly (Direct Citation), how often classes were cited together by the same patents (Co-Citation), how similar were the patterns of citations classes made or received from all other classes (Cosine Similarity, Inputs and Cosine Similarity, Outputs) and how similar were the patterns of citations classes made or received from all other patents (Cosine Similarity, Inputs, High Resolution and Cosine Similarity, Outputs, High Resolution). These different measures have typically been used to capture different aspects of technological proximity, as described above. However, all of the measures rely on patents' citations, and citations are determined by more than technologies' true proximity.

To clean the empirical signal of technology proximity from possible spurious relationships caused by the impinging factors, we compared the empirical proximity values to a null hypothesis: What would the measured proximity be by chance, given all the impinging factors? We calculated the random expectation by creating 1,000 randomized versions of the patent citation history, in which all of the impinging factors were exactly preserved. To create these randomized controls we identified groups of citations in which all the following properties were the same: the year the citing patents were issued, the year the cited patents were issued, and whether the citing and cited patents were in the same class (cross-class vs. same-class citations). For same-class citations, we only created citation groups in which all patents were in the same class. We then shuffled the cited patents among the citations in the group (Fig. \ref{citation_rewiring_diagram}). Perhaps surprisingly, virtually all citations were able to be grouped with other, similar citations and shuffled in this way (\textit{Appendix} \ref{Randomization_effectiveness}). The resulting shuffled versions of the network were thus different from the original, but preserved all the desired features of the number of patents in each class, the patent age sequence, etc. 

\begin{figure}[ht!]
\begin{center}
\includegraphics[width=\columnwidth]{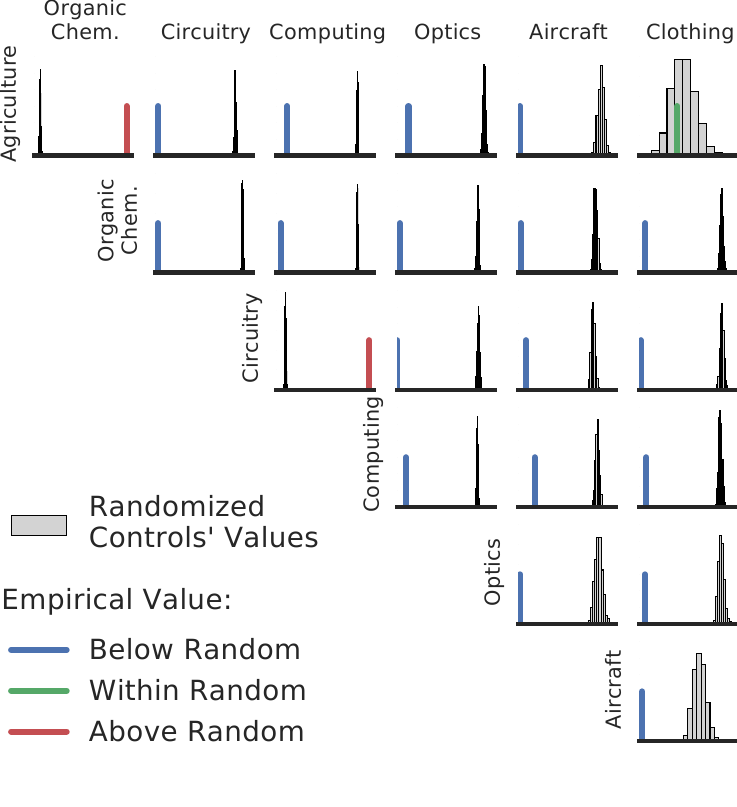} 
\end{center}
\caption{\textbf{The empirical value of proximity between any two classes was compared to the distribution of proximity values across 1,000 randomized controls for that link.} Each panel represents the link between two classes, the row class (e.g. Agriculture) and the column class (e.g. Clothing). The proximity metric shown is Co-Citation, the number of patents that cited patents in both the row class and the column class. Histogram: the distribution of proximity values for that link across 1,000 randomized controls. Vertical lines: the empirical value for that link derived from the original patent citation network, colored by whether it is below, within, or above the values of the randomized controls. The empirical proximity value for a given link was typically completely outside the distribution of the randomized controls' proximity values for that link (see Fig. \ref{related_unrelated_percentages}). The panels shown are a small sample of the 7,260 possible pairs of classes in the network of 121 IPC classes.}\label{rewiring_histogram}
\end{figure}

We used the randomized patent citation history to calculate the different proximity measures between the technology classes. For each pair of classes, we obtained a histogram of measured proximity values across the 1,000 randomized controls (Fig. \ref{rewiring_histogram}, gray bars). For certain measures and conditions it is also possible to calculate the complete probability distribution for the randomized controls, using analytic approximations (\textit{Appendix} \ref{analytic_approximations}). In contrast, the numerically-generated randomized controls are valid across all conditions and all measures.

We compared the histogram of proximity values from the randomized controls to the proximity value calculated from the empirical patent citation history (Fig. \ref{rewiring_histogram}, vertical lines). For most pairs of classes the empirical proximity measure was different from all 1,000 of the randomized control values, sitting entirely outside the histogram (Fig. \ref{rewiring_histogram}, (blue and red lines). This is analogous to the empirical link having a p-value below 0.001.

\subsection{Co-Classification and Co-Occurrence}
We controlled for the impinging factors in Co-Classification and Co-Occurrence measures by again comparing the empirical data to randomized controls. We created randomized versions of the patent record in which the number of associations made by each class, patent, inventor and firm were preserved. We also preserved temporal effects by treating each year of patent data separately, randomizing each individually, then combining them into a single, randomized version of history. We created 1,000 randomized controls in this way, described in more detail in \textit{Appendix} \ref{Cooccurrence_Methods}. We again compared the proximity measures calculated from the randomized controls to those of the empirical values, as with the citations in Fig. \ref{rewiring_histogram}.

\begin{figure*}[htbp!]
\begin{center}
\includegraphics[width=.8\textwidth]{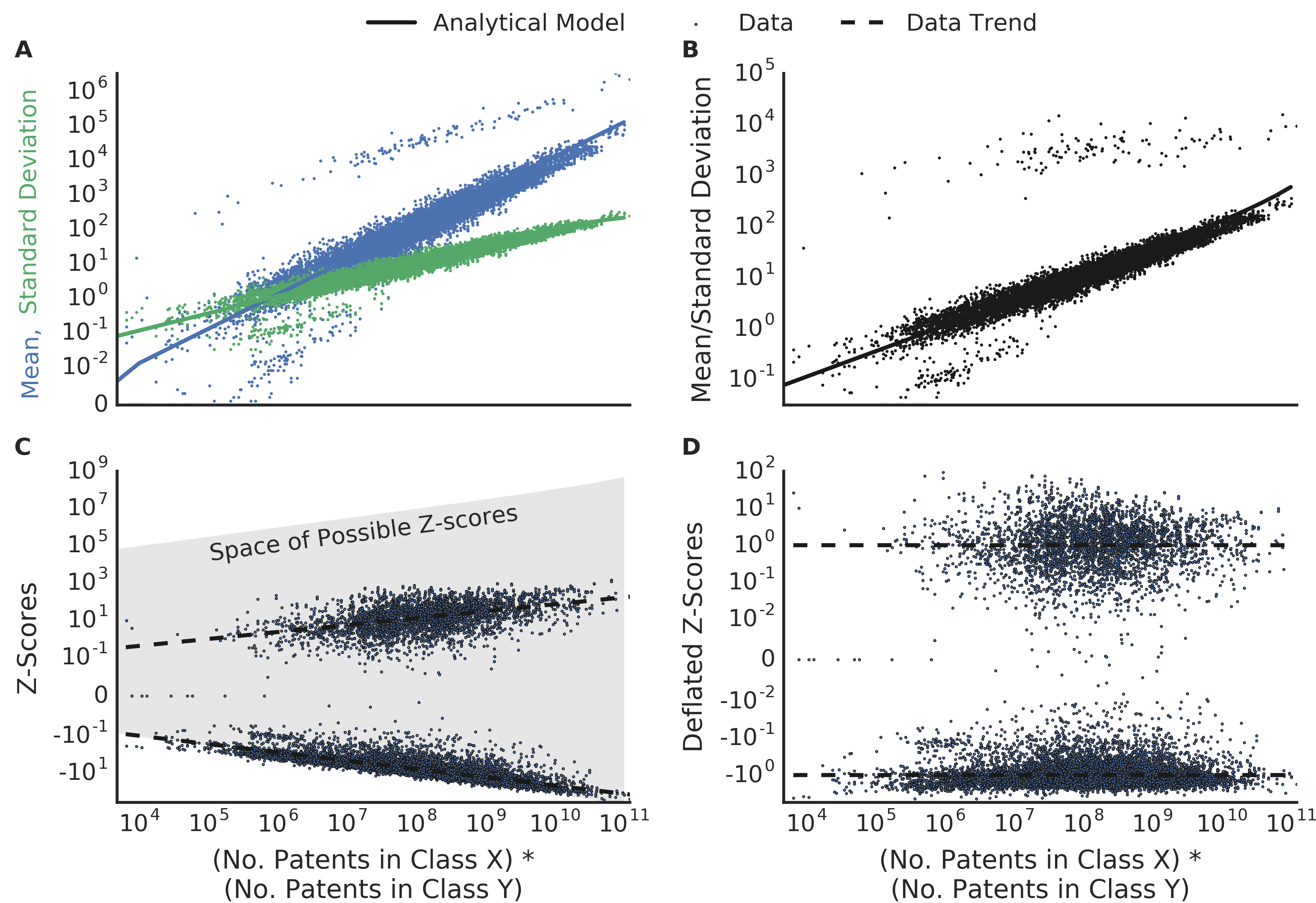} 
\end{center}
\caption{\textbf{The empirical proximity between a pair of classes can be compared to that of randomized controls as z-scores, but z-scores inflate with the number of patents in the classes, which requires a correction.} 
Solid lines: analytic model, dots: data, dashed lines: data trend.
A) The randomized controls' links' means and standard deviations, which both grew as the number of patents in the pair of classes grew, though at different rates.
B) The ratio of the randomized controls' links' means and standard deviations, which grew large as the number of patents in the pair of classes grows.
C) The empirical proximity values, compared to randomized controls using z-scores. The space of possible z-scores was limited above by the case where all patents in class $X$ cited all patents in class $Y$, and below by the case where none of the patents in class $X$ cited any patents in class $Y$. The limits of this space were modeled analytically (gray region), and the observed z-scores did grow in magnitude as the space expanded.
D) The z-scores were deflated by finding the inflation trends for the positive and negative z-scores, then dividing each z-score by the value predicted by the trend. The resulting deflated z-scores had no relationship with the number of patents in each class.
}\label{z-score_deflation}
\end{figure*}

\subsection{A Normalized Measure of Technology Proximity}
For each pair of classes, we summarized the difference between the empirical proximity and the random expectation through a z-score. From the  numerically-obtained distribution of randomized controls we calculated the mean ($\mu_{random}$) and the standard deviation ($\sigma_{random}$), then expressed the empirical proximity measure ($x_{empirical}$) as a z-score: $z = (x_{empirical} - \mu_{random})/\sigma_{random}$. The z-score expressed how more or less connected the pair of classes were than would be expected by chance, given the impinging factors.

The z-score values were nearly a completely normalized proximity measure, but z-scores are still affected by an impinging factor: the number of patents in each class. The maximum and minimum possible z-score for a link between two classes grow as the number of patents in each class grows; the space of valid z-scores increases and their magnitudes ``inflate" (Fig. \ref{z-score_deflation}). As an example, consider the Direct Citation measure between two classes $X$ and $Y$. All else equal, the number of expected citations between the two classes is determined by the number of patents in each class, $X_n$ and $Y_n$; the mean of the randomized controls grows proportionally with $X_n*Y_n$ (Fig. \ref{z-score_deflation}A, blue). However, the standard deviation of the randomized controls grows with the square root of $X_n*Y_n$ (Fig. \ref{z-score_deflation}A, green), and so the mean grows large relative to the standard deviation (Fig. \ref{z-score_deflation}B). The maximum distance an empirical value can be from the mean is also proportional to $X_n*Y_n$ (Fig. \ref{z-score_deflation}C, gray region): the minimum empirical value is always 0, and so the largest negative deviation from the mean is just the value of the mean; the maximum number of possible citations from $X$ to $Y$ is if every patent in $X$ cited every patent in the other, which is $X_n*Y_n$ itself and thus clearly proportional to $X_n*Y_n$. The links between large classes thus had a larger space of possible z-scores they could have, and indeed the observed z-scores grew in magnitude as the number of patents increased (Fig. \ref{z-score_deflation}C, dots).

The inflation of z-scores is not an empirical result, but an analytical relationship. It arises from three steps:
\begin{enumerate}
\item Randomized controls' mean value and standard deviation grow at different rates as the number of patents in a pair of classes increases, so their ratio increases.
\item The difference between the empirical value and the randomized controls' mean value (the numerator of the z-score) has a maximum possible value, and that value grows proportionally with the number of patents in both classes. 
\item As the maximum value of the numerator of the z-score grows, the denominator (the standard deviation) grows less quickly. The numerator grows large relative to the denominator, larger z-scores are possible, and the space of possible z-score values inflates.
\end{enumerate}

To illustrate that z-score inflation is definitional, we estimated analytically how the means and standard deviations of the randomized controls' link values would grow with $X_n*Y_n$ (Fig. \ref{z-score_deflation}A, B, solid lines; \ref{z-score_deflation}C, gray region). The analytic model is only illustrative, as the statistics of the randomized controls are due to a variety of factors and could thus deviate from the simple model. This is why some of the z-scores in Fig. \ref{z-score_deflation}C are actually below the lower barrier modeled, outside the gray region; the randomized controls' standard deviation was smaller expected, so the z-score's magnitude was larger than would otherwise be possible.

We corrected for the z-score inflation with a simple heuristic: regressing out the inflation trend. We calculated the trend of the z-score inflation for the positive and negative z-scores, then divided the z-score values by the trend line (Fig. \ref{z-score_deflation}C, D, dashed lines). We calculated the trend and regressed it out for each measure of proximity individually (Figures in \textit{Supporting Information}). The z-scores were thus deflated, and so the normalized measures of proximity had no correlation with class' number of patents.

The deflated z-scores were the desired measure of technological  proximity that was normalized to remove impinging factors. Empirical and normalized proximity measures could convey very different perspectives. For example, the empirical number of citations from ``Medical \& Hygiene" to ``Electric Communication" seemed large at $17,542$ citations, which put it in the top 3\% of citations. However, this was actually fewer citations than would be expected by chance: the randomized histories had $63,826 \pm 161$ citations between the two classes, producing a z-score of $-288.32$. After z-score deflation the normalized proximity was $-2.56$, in the bottom 3\% of normalized values.

\subsection{Robustness of Analysis to Different Classification Systems}
All patents were classified by patent agents under two systems: the United States Classification system (USPC: 430 classes) and the International Patent Classification system (low-resolution IPC3: 121 classes, high-resolution IPC4: 629 sub-classes). Data of patents’ main classes for all classification systems were available for patents from 1976-2010, and data from those years are used for all the relevant proximity measures. USPC secondary classification was available for patents from 1976-2010, and so Co-Classification using USPC reflect the same years of data as the other measures. For IPC3 and IPC4 data on secondary classifications was only available for patents from 1976-2006, and so the Co-Classification measure using these classifications reflect 4 fewer years of data.

The US patent office recently joined other national patent offices to exclusively use the new Cooperative Patent Classification system (CPC), which is based on the IPC. In order to ensure our findings are the most relevant for the future, we focus here on results from the IPC3, which is more similar to the modern CPC system than USPC and is more easily visualized than IPC4. We also repeated the analysis using the USPC and the higher-resolution IPC4 classification systems and found qualitatively similar results to those shown here (\textit{Appendix}).

\subsection{Data and Code for Reproduction and Extension of these Methods}
All code to perform these analyses and produce these figures is included online at \href{https://www.github.com/jeffalstott/technologyspace}. This code takes as input a set of raw data files describing patents’ classifications, authors, assignees and citations. These raw data files are in \textit{Supporting Information}. The code is written in Python and includes documentation.

\section{Results}

\subsection{Normalization Creates Closer Correlation of Different Measures of Technology Proximity}

\begin{figure}[]
\begin{center}
\includegraphics[width=\columnwidth]{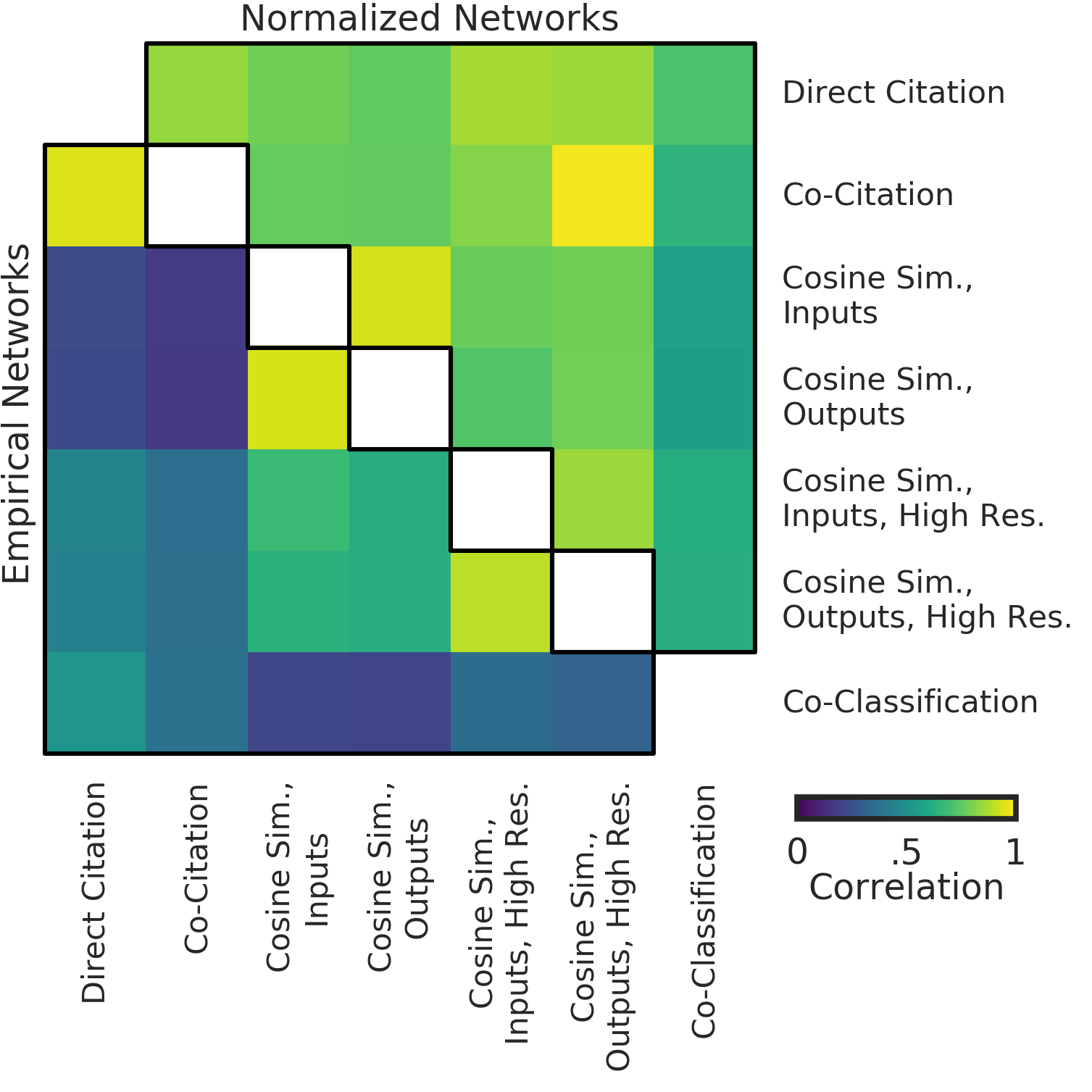}
\end{center}
\caption{\textbf{Normalizing networks made technology proximity networks more similar to each other, while less influenced by other factors.} Correlations are between the link weights of technology networks created with different measures of proximity. 
Lower left) Empirical networks.
Upper right) Normalized networks. Normalized networks' link weights are z-scores, where a link value of the empirical network is expressed as a z-score of the randomized controls' values for that link. Scatter plots and heat maps of the raw data for all comparisons of proximity measures are in Figs. \ref{Measure_Comparison_Scatter_Plots} and \ref{Measure_Comparison_Hex_Plots}.
}\label{measure_correlations}
\end{figure}

\begin{figure}[p!]
\centering
\includegraphics[width=\columnwidth]{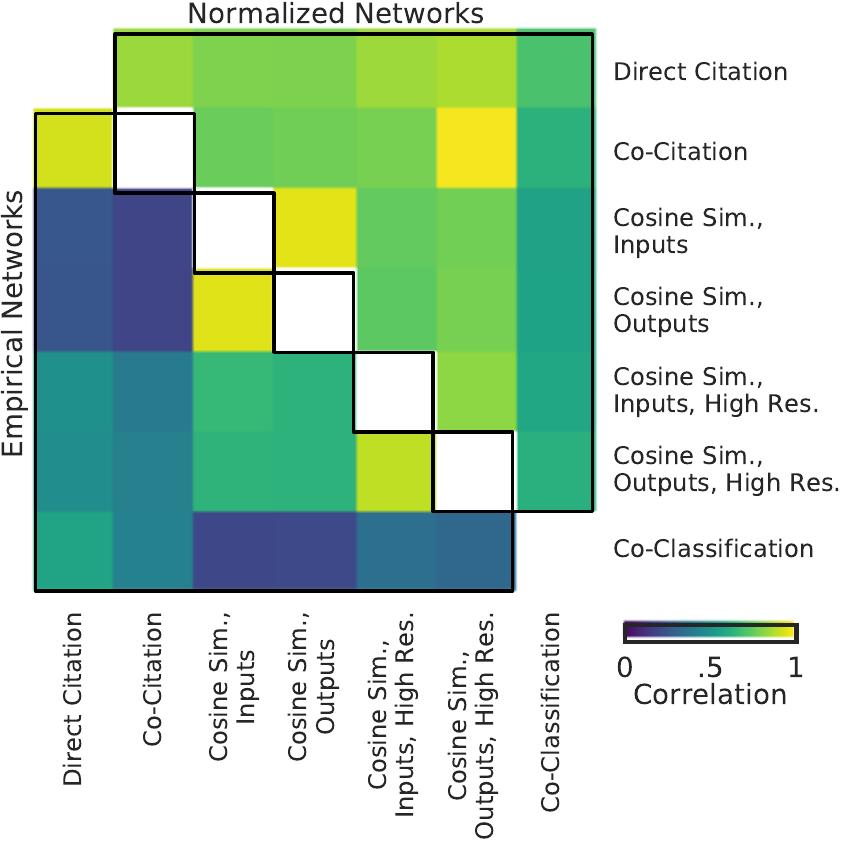} 
\caption{\textbf{The different measures of technology proximity, as calculated using the IPC3 classification system with patents from 1976-2006 for all measures.}}\label{Network_Correlations_Linear_2006_IPC}
\end{figure}

\begin{figure}[p!]
\centering
\includegraphics[width=\columnwidth]{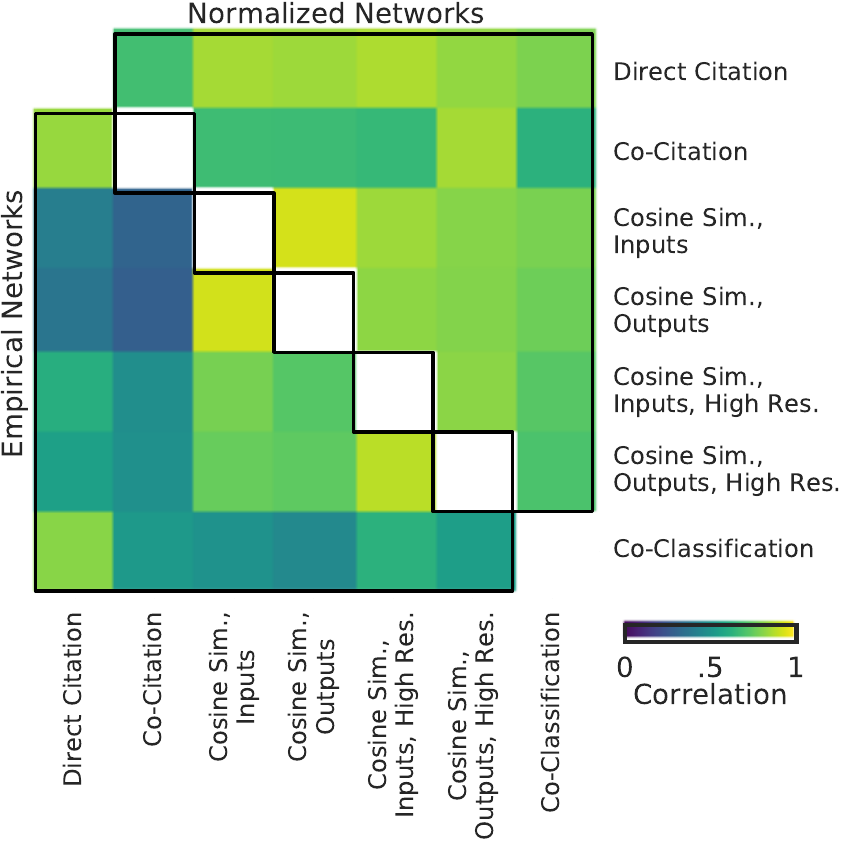} 
\caption{\textbf{The different measures of technology proximity, as calculated using the USPC classification system, had heterogeneous correlations before normalization. After normalization, however, all measures correlated.}}\label{Network_Correlations_Linear_USPC}
\end{figure}

\begin{figure}[p!]
\centering
\includegraphics[width=\columnwidth]{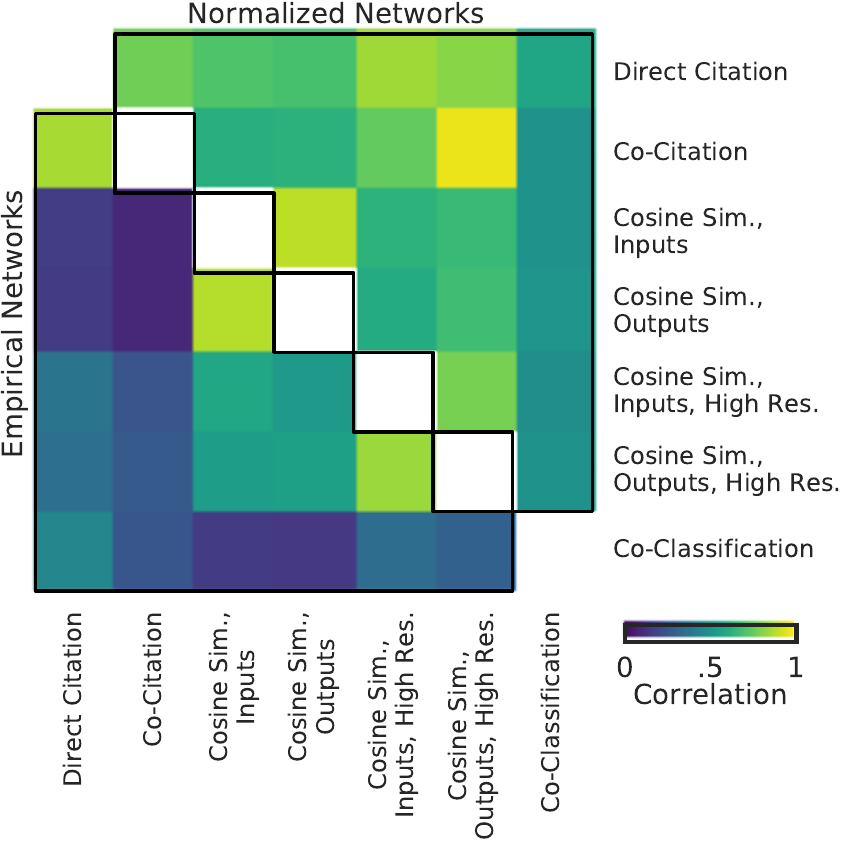} 
\caption{\textbf{The different measures of technology proximity, as calculated using the IPC4 classification system, had heterogeneous correlations before normalization. After normalization, however, all measures correlated.}}\label{Network_Correlations_Linear_IPC4}
\end{figure}

Normalization changed how the different kinds of proximity measures compared to each other. Among the empirical networks (before normalization) there were three groups of correlated networks (Fig. \ref{measure_correlations}, lower left panel). In the first group were Direct Citation and Co-Citation, in the second were the four varieties of Cosine Similarity, and last was Co\hyp{}Classification.
After normalization, all of the proximity measures became more correlated with each other measure (Fig. \ref{measure_correlations}, upper right panel). For classification with IPC3 (Fig. \ref{measure_correlations}) and IPC4 (Fig. \ref{Network_Correlations_Linear_IPC4}), Co-Classification was less correlated with other measures, though its correlation was also raised after normalization\footnote{The Co-Classification network for IPC3 and IPC4 only included the available patent data from 1976 to 2010, unlike the other proximity measures. However, similarly restricting the other proximity measures to data only from 1976 to 2006 shows the same lower correlation for Co-Classification, and the same overall correlation structure across all measures (Fig. \ref{Network_Correlations_Linear_2006_IPC}).}. In contrast, Co-Classification was as correlated as other measures when using classification with USPC; the less-correlated measure with USPC was Co-Citation (Fig. \ref{Network_Correlations_Linear_USPC}).

Thus, removing impinging factors led to more agreement among the different measures of proximity. There is little \textit{a priori} reason to expect that normalization would lead to increased agreement, though we discuss possible reasons below. However, \textit{a posteriori} the increased agreement of different proximity measures validates the utility of the method of normalizing out impinging factors.

\subsection{The Technology Network is Sparse}
\begin{figure}[b]
\begin{center}
\includegraphics[width=\columnwidth]{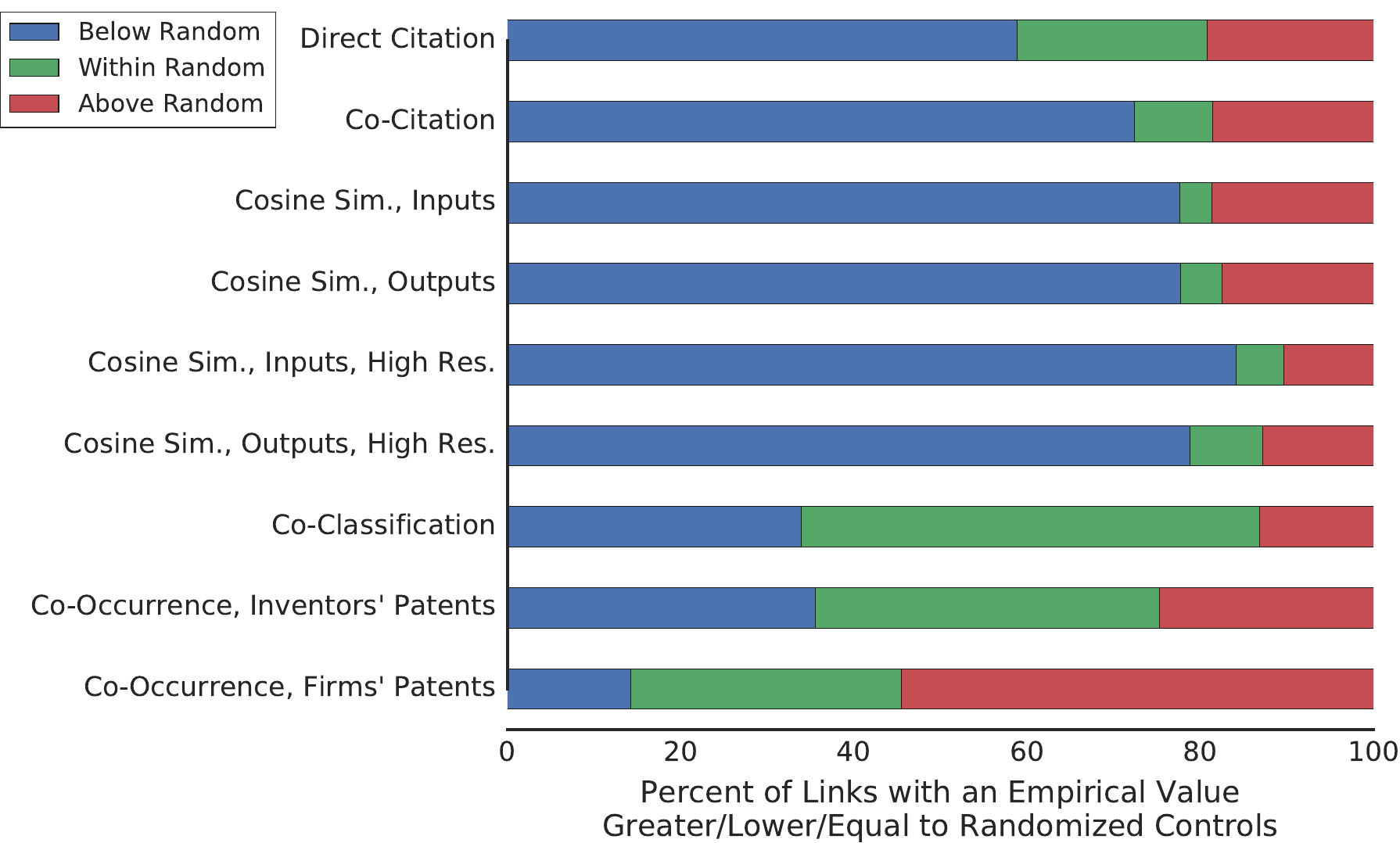} 
\end{center}
\caption{\textbf{Few links between classes had a higher proximity than that expected by chance.} Technology networks created using different measres were compared to 1,000 randomized controls, by comparing the weights of their links. For most networks, a majority of the empirical links had lower proximity than any of the randomized controls (blue), and a minority were above the randomized controls (red). The exception was Co-Occurrence, Firm networks, for which many more pairs of technology classes had greater Co-Occurrence than would be expected by chance.
}\label{related_unrelated_percentages}
\end{figure}

Before normalizing proximity measures, it is difficult to quantify the distance between two technologies and then assert if the resulting number is a high or low value. However, comparing an empirical link weight to randomized controls yields a natural interpretation for whether a link is particularly strong or weak: an empirical link weight is stronger or weaker \textit{than would be expected by chance.} Most links in the networks were weaker than would be expected by chance (Fig. \ref{related_unrelated_percentages}), measured as being below the link weights of any of the 1,000 randomized controls (analogous to $p<.001$). For most proximity measures, \textasciitilde{}20\% of the links were stronger than chance, indicating the two technology classes they connected were particularly related.

\begin{figure}[p!]
\centering
\includegraphics[width=\columnwidth]{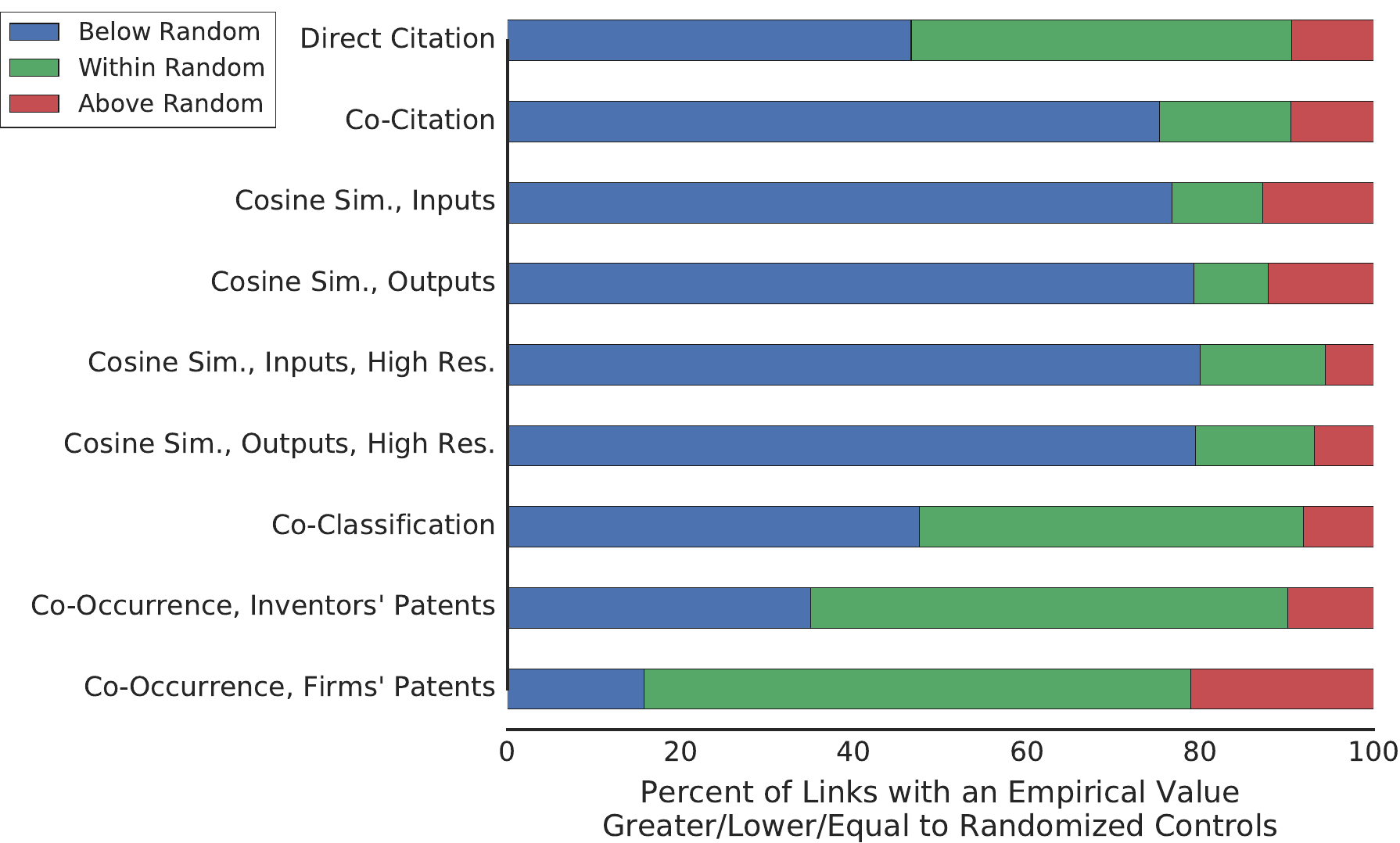} 
\caption{\textbf{Using the USPC classification system, all measures of technology proximity showed a sparse network after normalization.}
}\label{related_unrelated_percentage_USPC}
\end{figure}

\begin{figure}[p!]
\centering
\includegraphics[width=\columnwidth]{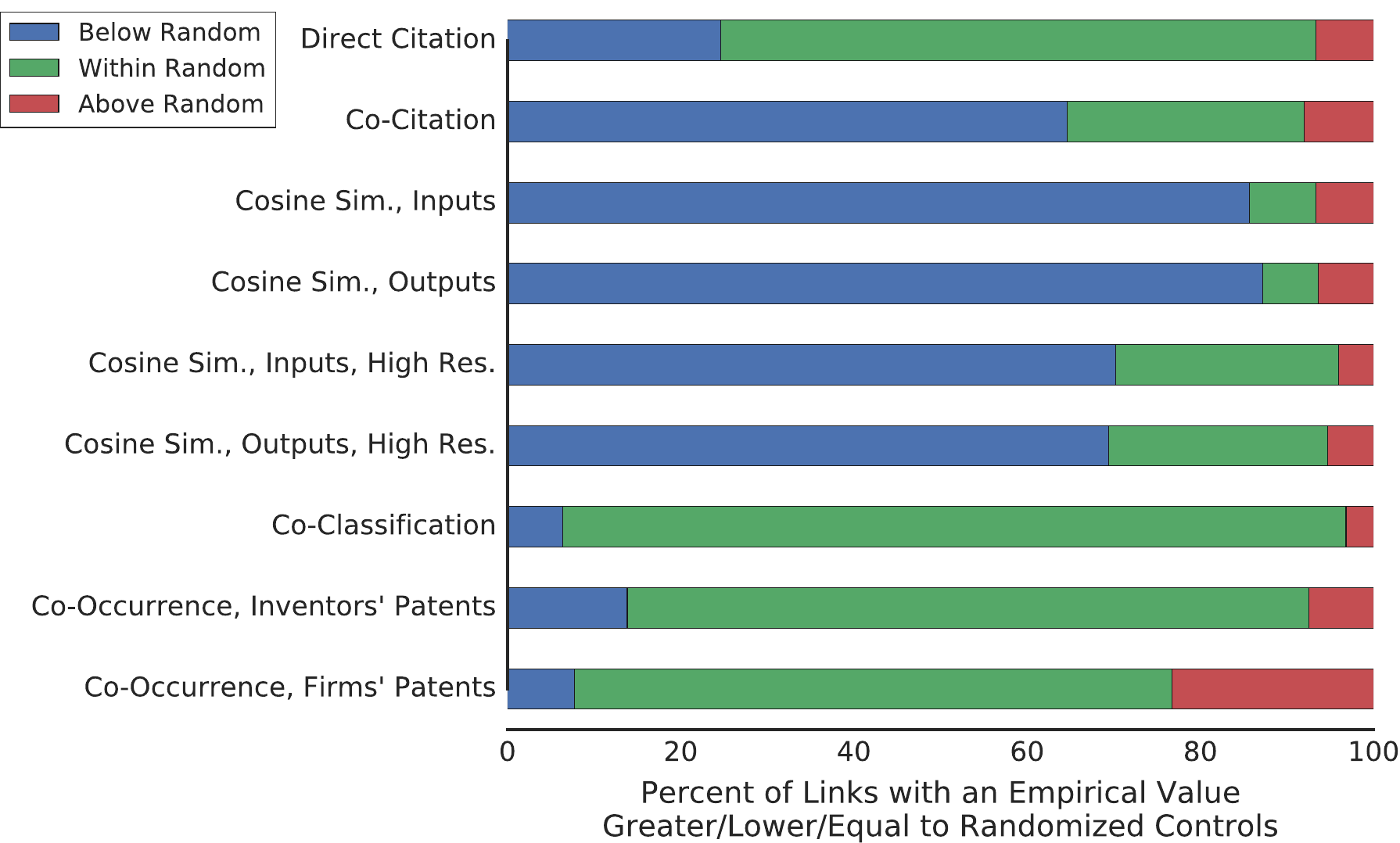} 
\caption{\textbf{Using the IPC4 classification system, all measures of technology proximity showed a sparse network after normalization.}}\label{related_unrelated_percentage_IPC4}
\end{figure}

The technology networks were thus sparsely connected, though sparse is a relative term; the network had 7,260 possible links, and with a link density 20\% there were still 1,452 links remaining. However, using other classification systems with higher resolution revealed an even sparser network: using the USPC classification (430 classes) yielded a sparsity of \textasciitilde{}10\%, and using the IPC4 classification (629 classes) yielded a sparsity of \textasciitilde{}5\% (Figs. \ref{related_unrelated_percentage_USPC}, \ref{related_unrelated_percentage_IPC4}).

Controlling for spurious factors quantifies an intuitive fact: most technologies are not particularly proximate to each other. Instead, any single technology is only notably proximate to a fraction of the other technology domains. This finding on the technology map structure again justifies the value of our method of normalizing out impinging factors.

Sparsity was notably less the case for the Co-Occurrence measured from firms' patenting histories: technology pairs occurred together in firms' patenting histories at rates greater than chance about twice as often as in the other networks. Using the IPC3 classification system this meant \textit{most} technology pairs were significantly connected. However, using the higher resolution USPC or IPC4 classifications lowered the frequency of significant technology pairs to \textasciitilde{}20\% (Figs. \ref{related_unrelated_percentage_USPC}, \ref{related_unrelated_percentage_IPC4}).

We also analyzed the patenting histories of countries and found a similar pattern to that of firms, though the comparatively small sample of $<200$ countries meant the trend was not statistically significant (\textit{Appendix} \ref{cooccurrence_country}). 

\subsection{Inventors' Behavior Follows Proximity Measures Closely, While Firms' Portfolios Follow Less Closely}
\begin{figure}[tbp]
\begin{center}
\includegraphics[width=\columnwidth]{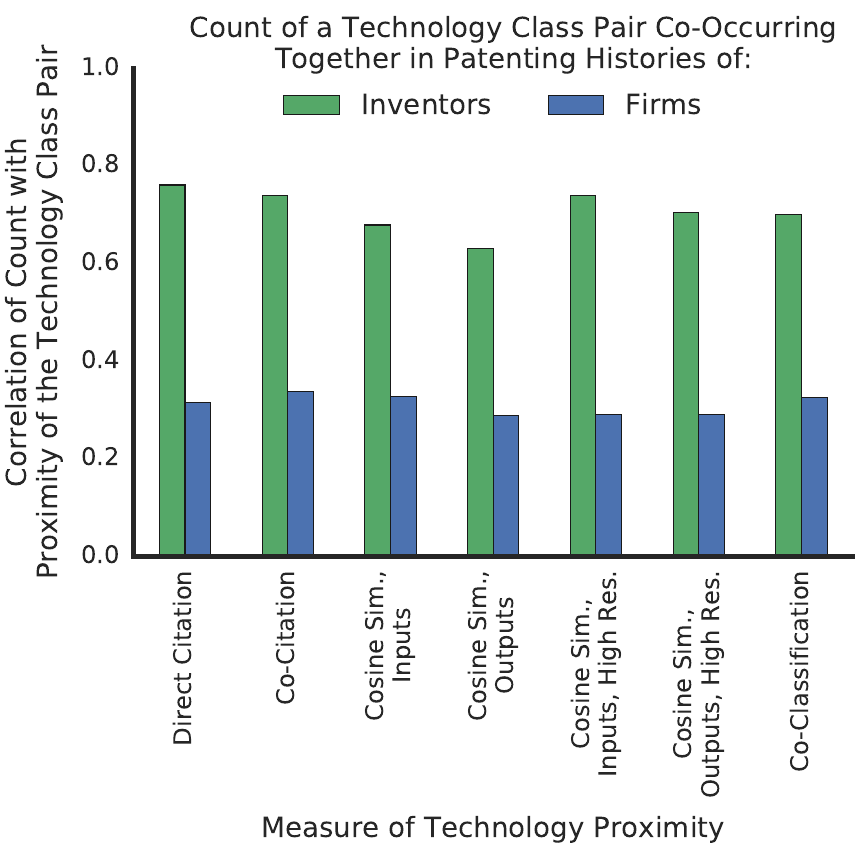} 
\end{center}
\caption{\textbf{Inventors closely followed the technology network maps derived from the normalized proximity measures, while firms followed the maps less closely.} The normalized counts of how often two technology classes co-occurred in inventors' patenting history correlated with other normalized measures of technology proximity (blue bars). The normalized counts of how often two technology classes co-occurred in firms' patent portfolios correlated only modestly with the other measures of proximity (green bars).
}\label{Relatedness_Behavior_Correlation}
\end{figure}

Inventors' patenting histories closely followed the technology network structure identified by the normalized measures. Pairs of technology classes' normalized rates of Co-Occurrence in inventors' patent histories were strongly correlated with the other citation- and classification-based networks (Fig. \ref{Relatedness_Behavior_Correlation}, blue bars). The normalized technology networks, then, not only began to converge on a common description of technologies' proximity to each other, but to a description that also mirrored inventors' behavior. The technology network maps may thus provide explanation for why a single mind that is able to invent in ``organic chemistry" is also likely able to invent in ``agriculture": these technology domains are intrinsically related. Thus, this normalized map will be particularly useful for analyzing and predicting the invention portfolios and learning paths of inventors.

\begin{figure*}[]
\centering
\includegraphics[width=\textwidth]{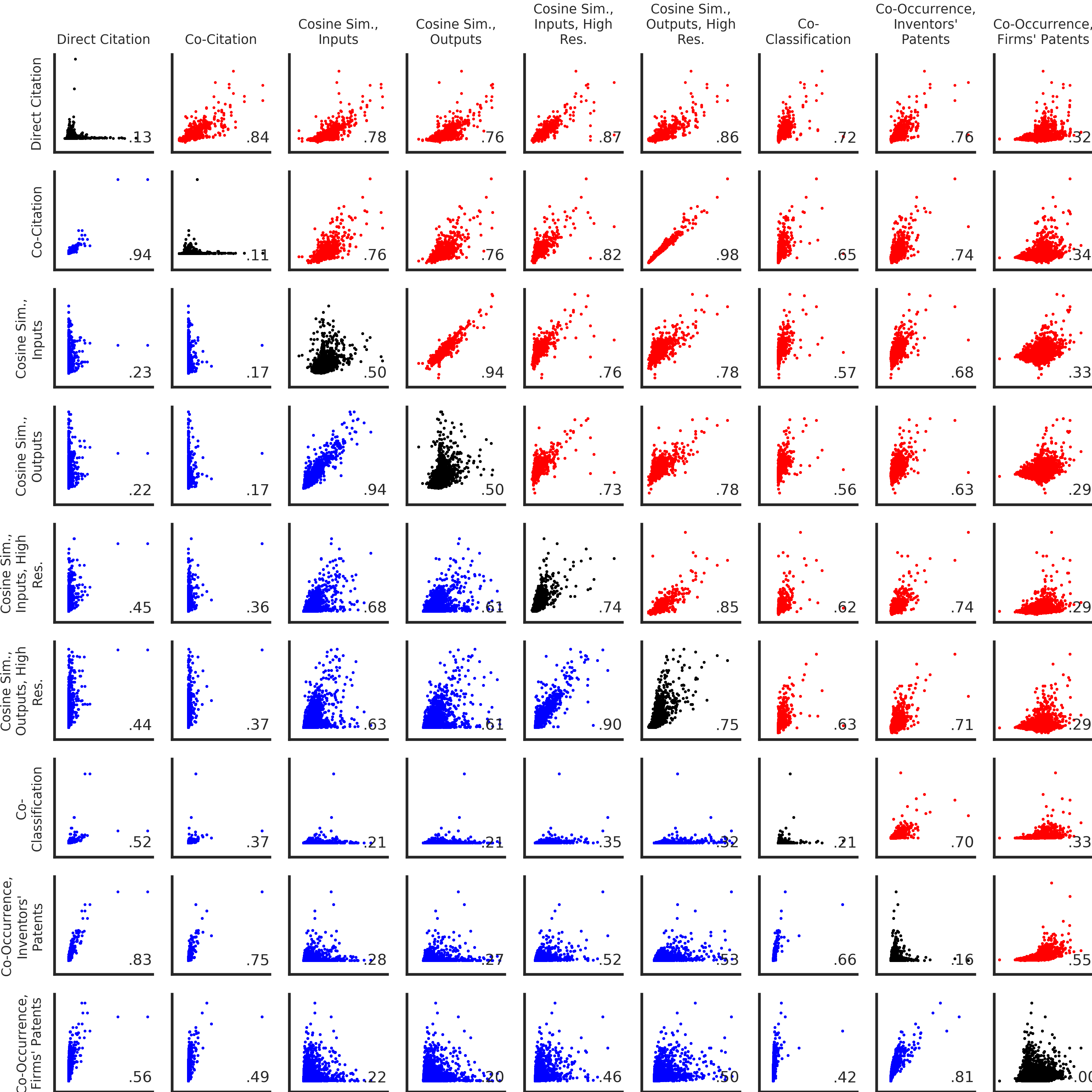} 
\caption{\textbf{Comparison of link weights across different proximity measures, as scatter plots.} Blue: empirical networks. Red: normalized networks. Black: empirical vs. normalized link weight for each measure compared to itself (x-axis: empirical, y-axis: normalized). Text: Pearson correlation.
}\label{Measure_Comparison_Scatter_Plots}
\end{figure*}

Firms' patent portfolios, in contrast, followed the other technology networks less closely. Pairs of technology classes' normalized rates of Co-Occurrence in firms' patent histories were also correlated with the other networks, but only modestly (Fig. \ref{Relatedness_Behavior_Correlation}, green bars). The association between technology proximity and inventive behavior is similar for firms and inventors, but the variance is much larger for firms, reducing the strength of the association (Fig. \ref{Measure_Comparison_Scatter_Plots}). Firms, then, are like inventors in that they tend to invent in classes related to those that they already have experience in; notwithstanding, deviations from this general pattern are much more common and sizable for firms. Previous research has used co-occurrence data to investigate whether firms preferentially diversify into related classes \cite{Bottazzi2010, Breschi2003, Teece1994}. The present results show that firms' patent portfolios are indeed influenced by the proximity of different technology domains, but this is just one, modest influence. Firms' decisions to enter into a new technology domain are also determined by such factors as market demand, the availability of capital and risk diversification. Furthermore, firms are less constrained than individual inventors; they can hire additional staff or acquire new ventures that can bring in new knowledge unrelated or dissimilar to a firm's previous capabilities. As such, firms' inventive behavior only partially reflects other measures of the technology proximity space. 

\subsection{Network Stability Over Time}
The technology networks presented so far were constructed using all available patent data from 1976 to 2010. These networks represent the ''view from 2010". How much does the network change if we consider the view from different points in history? We previously analyzed the temporal stability of the unnormalized technology proximity measures, and found that they altered little over time\cite{Yan2015}. Here we analyzed the temporal stability of the normalized technology networks by constructing multiple versions using data up year $X$, where $X$ is each year from 1976 to 2010. We assessed the similarity of the networks over time, again by measuring the correlation of their link weights. We measured the correlation of a network created with data from each year X with the network created with data from year $X-1$.

\begin{figure*}[htbp!]
\centering
\includegraphics[width=.8\textwidth]{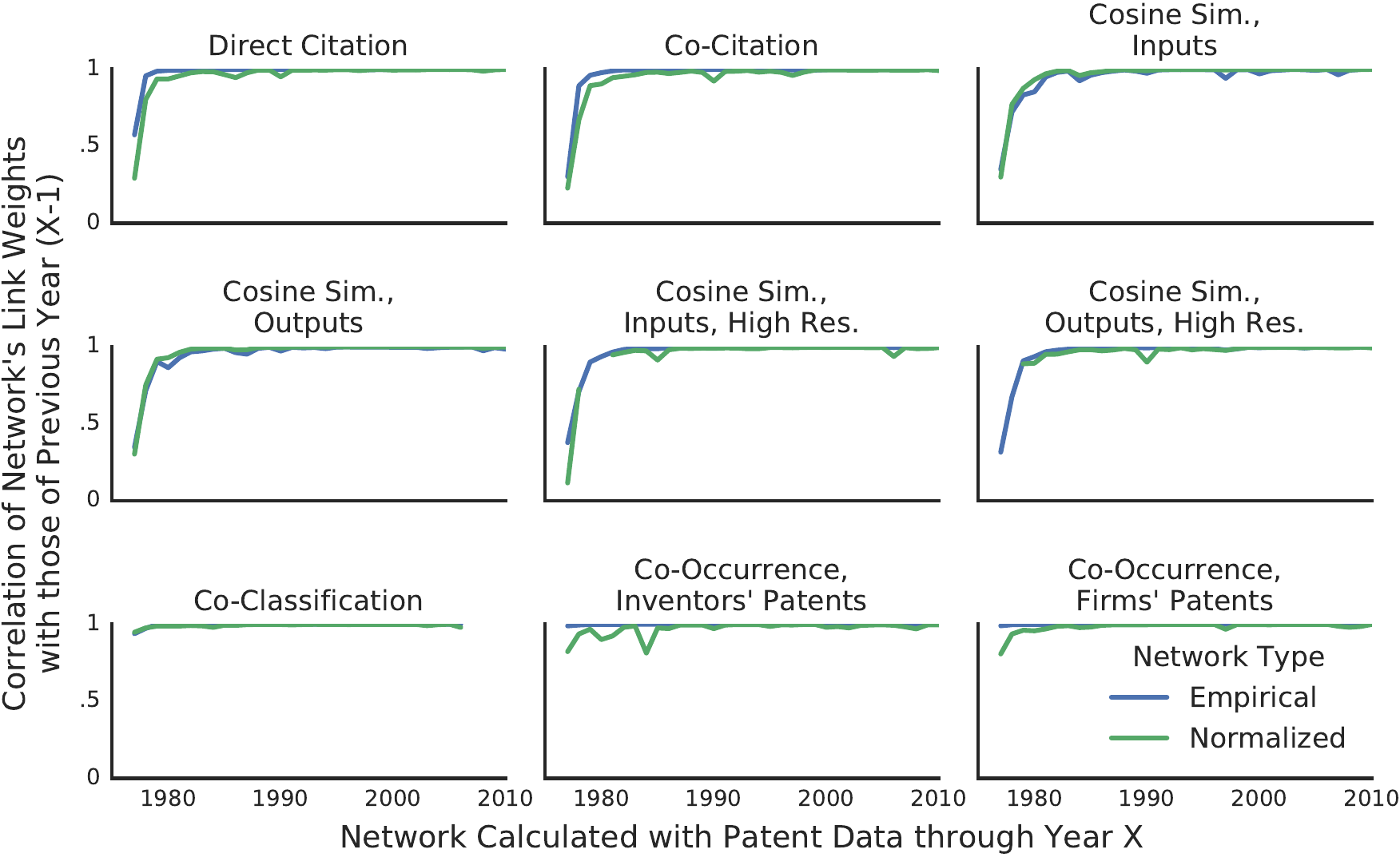} 
\caption{\textbf{All measures of technology proximity were stable over time.} Lines: the correlation of the link weights of a technology proximity network, calculated with data from 1976 to year $X$, with the link weights of the same network calculated with data from 1976 to year $X-1$. Each panel corresponds to a different measure of technology proximity. Blue lines: the empirical technology proximity. Green lines: the normalized measure of technology proximity. 
}\label{Network_Temporal_Stability}
\end{figure*}

The networks started out unstable in the 1970s, but quickly became stable by the 1980s and remained so through 2010 (Fig. \ref{Network_Temporal_Stability}). The initial instability was likely due to comparatively little data in these first few years. With adequate data, however, the normalized networks are generally stable over time. Thus, the technology proximities calculated here are not particular to the year 2010, but reflect a more lasting set of relationships. These relationships may of course still change slowly or locally, and large changes can not be ruled out before 1976 or after 2010.

\section{Discussion} 
Normalizing technology network maps by controlling for impinging factors uncovers previously obscured information, such as the sparsity of the space of technology proximity. Normalization also leads to convergence of many of the different measures of proximity, as seen through their increased correlation. These results validate the method of normalizing proximity measures to remove impinging factors and the usefulness of resulting maps. 

\subsection{Technology Proximity, the Inventive Process, and Technology Development}
After normalization, different technology proximity measures began to provide a similar map of the technology space. There is no \textit{a priori} reason why this should happen. In fact, one could assume that proximity measures based on different perspectives should lead to different maps of the technology space. However, the increased agreement of the different measures could be explained by that the simple relationship of inventive inputs, inventive outputs, and inventive processes becomes more apparent after impinging factors and their spurious effects are removed.

Let us assume that inventions are the result of a cognitive process, $f()$, that transforms knowledge inputs $x$ into inventive outputs: $y=f(x)$. The following then holds: $f_1(x_1)=f_2(x_2′)$ if $f_1=f_2′$ and $x_1=x_2′$. Two identical inventive processes that are given the same inputs will yield the same outputs. We can then relax the concept of equality to just similarity: similar inventive processes that are given similar inputs will yield similar outputs. We can then explain why technology classes that frequently Co-Occur in inventor's portfolios also have similar sets of inputs or outputs: technology classes that require similar inventive processes (``functions") are being used on similar inputs to obtain similar outputs. Inventors with similar inventive process functions $f$ are using similar inputs $x$ and thus getting similar outputs $y$, or seeking to create similar outputs $y$ and thus are using similar inputs $x$. Different measures of technology proximity that measure input similarity vs. output similarity thus correlate, because inputs and outputs are linked by the function that turns inputs into outputs.

Using normalized measures of technology proximity, we now better understand the statistics of how inventors use their knowledge to create new inventions. The maps presented here are a cleaned signal of technology proximity, which we can now use to address further questions about how technology proximity affects technology development on a large scale. For instance, it has been hypothesized that the growth or decay of one technology domain is affected by technology dynamics and competition in neighboring domains \cite{Saviotti1995,Bruckner1994} and that technology development in one area benefits from knowledge spillovers induced by R\&D investments in proximate technologies \cite{Verspagen1997,Verspagen1997a,Bloom2013}. The normalized network maps can be used to test these hypotheses, and so may improve our ability to explain specific classes’ number of patents, or how proximate technology domains interact and even compete. If the network maps can be used to explain domains’ dynamic coupling, then the maps could be used for prediction: if a new development arises in one domain, we could predict cascading effects reaching into proximate domains.

\subsection{Absolutely Related vs. Particularly Related}
It is important to highlight that the methods introduced here do not measure if two technological domains are proximate in an absolute sense; they only measure if two classes are particularly proximate. As an example, consider a class with just one patent that is super-connected, which cites every other patent. Randomizing this citation history while also preserving each patents' outward and inward citations would yield the exact same arrangement: the super-connected patent would still cite every other patent. The randomized controls would not be able to deviate from the observed data, because only the original arrangement of citations satisfies the conditions. The empirical links between the super-connected class and other classes would look exactly the same as the randomized controls' links, and so the z-scores for those links would be 0. The class with the super-connected patent could thus reasonably be considered very proximate to \textit{every} other class, but the methods we introduce here would just see it as not particularly proximate to \textit{any} other class. This example illustrates that we do not measure if two technology classes are proximate in an \textit{absolute} sense, but only if, given their absolute level of connection to all other classes, they have \textit{particular} connectivity or disconnectivity to each other above what could be expected by chance.

\subsection{Alternative Definition of Impinging Factors}
We identified several aspects of the patent record as not pertaining to the proximity of technological artifacts, arising instead through the patenting or inventive process. We called these aspects of the patent record impinging factors, because they affect measures of technological proximity without contributing to the true signal of proximity. These impinging factors included patents' number of citations, the age of the cited patents, the number of technology classes an inventor or firm has patented in, etc.; we described above our assessments of what drives these factors. However, the methods that we present here could be modified to control for additional impinging factors, such as whether inventors' patents' preferentially cite patents by other inventors in the same city or firm. It is also possible to control for fewer factors, if a researcher wanted those factors to affect the measure of proximity. The general method simply requires clearly identifying which properties of the patent record are to be controlled for, then generating randomized controls that preserve those properties.

\subsection{Utility of the Maps and Mapping Technique}
The normalized technology networks measured here can now serve as map for uses in technology development planning and management. Both individual inventors and firms can locate themselves and their knowledge on the map and observe what technology domains are nearby in the technology space. Nearby domains are likely easier targets for new invention over more distant domains. Inventors are particularly justified in using the map to guide their future inventions, since inventors who successfully patent in multiple domains typically do so in proximate domains. Firms are less justified in restricting themselves to targeting technologies in solely those domains that are proximate to their existing knowledge base; they may instead hire additional inventors with new knowledge to roam further afield. For both inventors and firms, it may be possible to use the map to plan a long-term research and development path: starting in the domain where one currently has knowledge, one can target at a series of domains that are always proximate to each other, but ultimately result in patenting in a domain very unrelated from one's origin. Thus, the normalized technology proximity map can be a significant strategic planning tool.

Strategies of following the map (or not) are justified if an individual or firm wants to behave like those who successfully patent. It is possible, however, to have a higher bar: to be an inventor whose patents receive many citations, or to be a firm whose inventions yield high financial returns. Multiple lines of research have found that inventive efforts that combine or move between proximate technologies are more likely to successfully invent, but the results are low novelty and unlikely to be a breakthrough with high impact \cite{Fleming2001,Fleming2007, Simonton1999,Chan2011,Chan2015}. In contrast, inventive efforts that combine distant technologies are less likely to succeed in creating an invention, but if this hurdle is overcome then the results are more novel and more likely to be a breakthrough. The network maps presented here provide a cleaner measure of technology proximity to further test this theory: it is possible to identify inventors and organizations that persistently followed the map versus not, and then examine their performance. Analyzing inventive behavior in terms of a network may also reveal more complexity: high-performing inventors or firms may indeed follow the network more or less closely, but they may also employ more complex strategies like targeting particularly dense or sparse regions of the network.

Additionally, the normalization methods presented here are also of potential relevance in other domains, such as the study of the proximity of scientific fields. The same randomization approaches can be used to map the latent space of scientific disciplines using journal articles' citations, classifications, etc. Measuring the space of scientific topics with greater accuracy and statistical rigor may allow for answering such questions as whether the processes of intellectual creation follow universal rules, regardless of the scientific or technological nature of the knowledge involved.

\subsection{Conclusion}
Technology is a complex system, but we can gain understanding of that system by mapping out its components and their relations to each other. With the more accurate patent-based technology mapping techniques presented here it is possible to study technology development with a new level of clarity, including both aspects of technologies themselves and how humans interact with those technologies. Improved understanding of technologies and invention may ultimately inform better technology development policies, leading to more successful technology innovation and management.

\begin{acknowledgements}
JA, GT and JL designed the study. BY collected the data and visualized the networks. JA and GT developed and performed the analysis. JA, GT and JL wrote the paper. We thank Ulf Bissbort, Bikramjit Das, Tommaso Demarie, Francois Lafond, and Chris Magee for helpful discussions. This work was supported by the SUTD-MIT International Design Centre (IDG31300112) and by the Singapore Ministry of Education Tier 2 Academic Research Grants (T2MOE1403).
\end{acknowledgements}


\newpage
\begin{center}
\textbf{\large Appendix}
\end{center}

\section{Network Visualization}\label{network_visualization}
We visualized the empirical and normalized technology proximity networks
(Fig. \ref{Empirical_Normalized_Networks}A-B). We highlighted a community structure for each network, which was calculated by approximately maximizing the weighted modularity using a faster version of the Louvain method \cite{Newman2004, Blondel2008, Traag2015}. Only a subgraph of the networks' links were visualized: the planar maximally filtered graphs \cite{Tumminello2005}. These graphs contained the set of links with the highest weights that were also topologically planar, such that they could be laid out flat on a plane without links crossing.

\section{Effectiveness of the patent citation network randomization}\label{Randomization_effectiveness}
\begin{figure}[h]
\begin{center}
\includegraphics[width=\columnwidth]{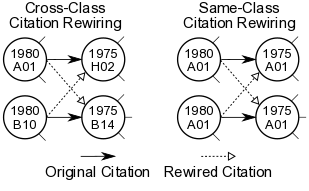} 
\end{center}
\caption{\textbf{The citation rewiring process used to create randomized control networks preserved many properties of the original patent citation network.} Citations were selected whose citing patents were issued in the same year and whose cited patents were also issued in the same year. Groups of citations were selected that were either all cross-class (left) or all same-class and all within the same class (right). The citations in the group were then shuffled. Performing this shuffling operation resulted in a randomized version of the patent citation network that still preserved many properties of the original networks, such as the cross-class citation rates, the time lag of citations, and the number of citations made and received by each patent. }\label{citation_rewiring_diagram}
\end{figure}

\begin{figure}[h]
\centering
\includegraphics[width=\columnwidth]{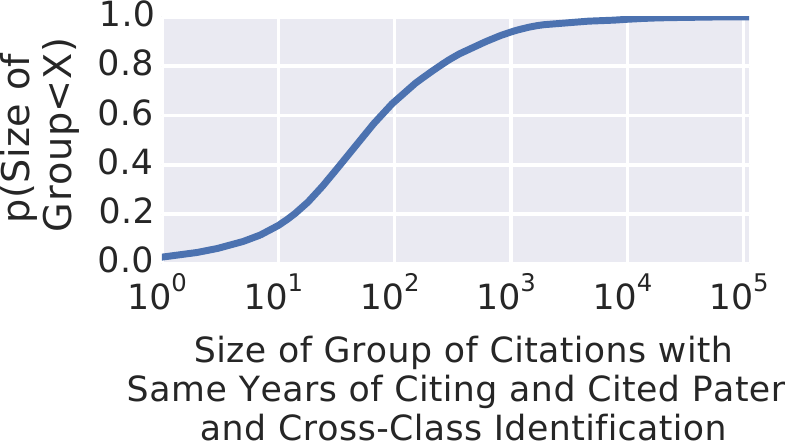} 
\caption{\textbf{Most citations had many other citations with the same properties, and so could be rewired.} The rewiring procedure used to create randomized controls required pairing each citation with another citation with similar properties (Fig. \ref{citation_rewiring_diagram}). If the group of citations with those properties was small, then that link would not have many opportunities for rewiring, and so the randomized controls would be similar to the empirical network. Over 85\% of citations were in a group that had 10 or more members. Only 2.55\% of citations were unique and could not be rewired. In practice, the empirical networks rarely resembled the randomized controls (Fig. \ref{related_unrelated_percentages})}\label{rewiring_groups}
\end{figure}

There are two wrinkles in how the generation of the randomized controls, which could in theory could affect the normalized proximity measures, though in practice do not. The first wrinkle is that it is not guaranteed that for every citation there is another citation to be paired with that has all the same properties required. Fortunately, this happened rarely. Each citation was part of group of citations that had the same citing patent year, cited patent year, and cross-class identification (and for same-class citations, being within a particular technology class). Fig. \ref{rewiring_groups} shows the number of citations in each group that was represented in the patent network. Only approximately 14.16\% of citations were part of a group that fewer than 10 members. 2.55\% of citations were part of a group with 1 member; these were unique citations, and could not be rewired. Leaving these citations unaltered made the randomized control networks more similar to the empirical network. As discussed in the main text, the randomized control networks and the empirical network were still very different, and so the effect of the unrewired links was unappreciable.

The second wrinkle of this normalization method is with rewiring cross-class citations. As in Fig. \ref{citation_rewiring_diagram}, if all four patents are from different classes, the desired outcome is achieved. However, it is possible that the citing patent of one citation is in the same class as the cited patent of the \textit{other} citation. In this case, both citations are indeed cross-class citations, but after rewiring one of the citations would become a same-class citation. Thus, the cross-class citation rate would decrease in the randomized controls, and the same-class citation rate would decrease (Note that it is not possible for the reverse mistake to occur, because all the same-class citations are paired and swapped separately.). The solution, of course, is to check that the paired citations do not have the problematic arrangement of classes, and so will not create a same-class citation. While this works in principle, it does not in practice. The solution requires checking, rejecting, and re-suggesting possible pairs of citations. This process creates significant computational problems, and it is hard to assess if and when the process will even converge. Because of this, we left the problem in place, and so the randomized controls had an increased rate of same-class citations. This rate was small, with the rate of same-class citations raising from 39.74\% in the empirical patent citation network to a typical rate of 41\% in randomized controls. The decrease in cross-class citations in the randomized control patent citation networks would typically make the empirical network appear to have an unusually high amount of citations between two classes, leading to an unusually strong connection in the technology network. However, as discussed in the main text, the empirical network generally had much lower proximity values than would be expected by chance. As such, the error that the imperfect rewiring introduces would only make the unusually low normalized values more notable.

\begin{figure*}[]
\centering
\includegraphics[width=\textwidth]{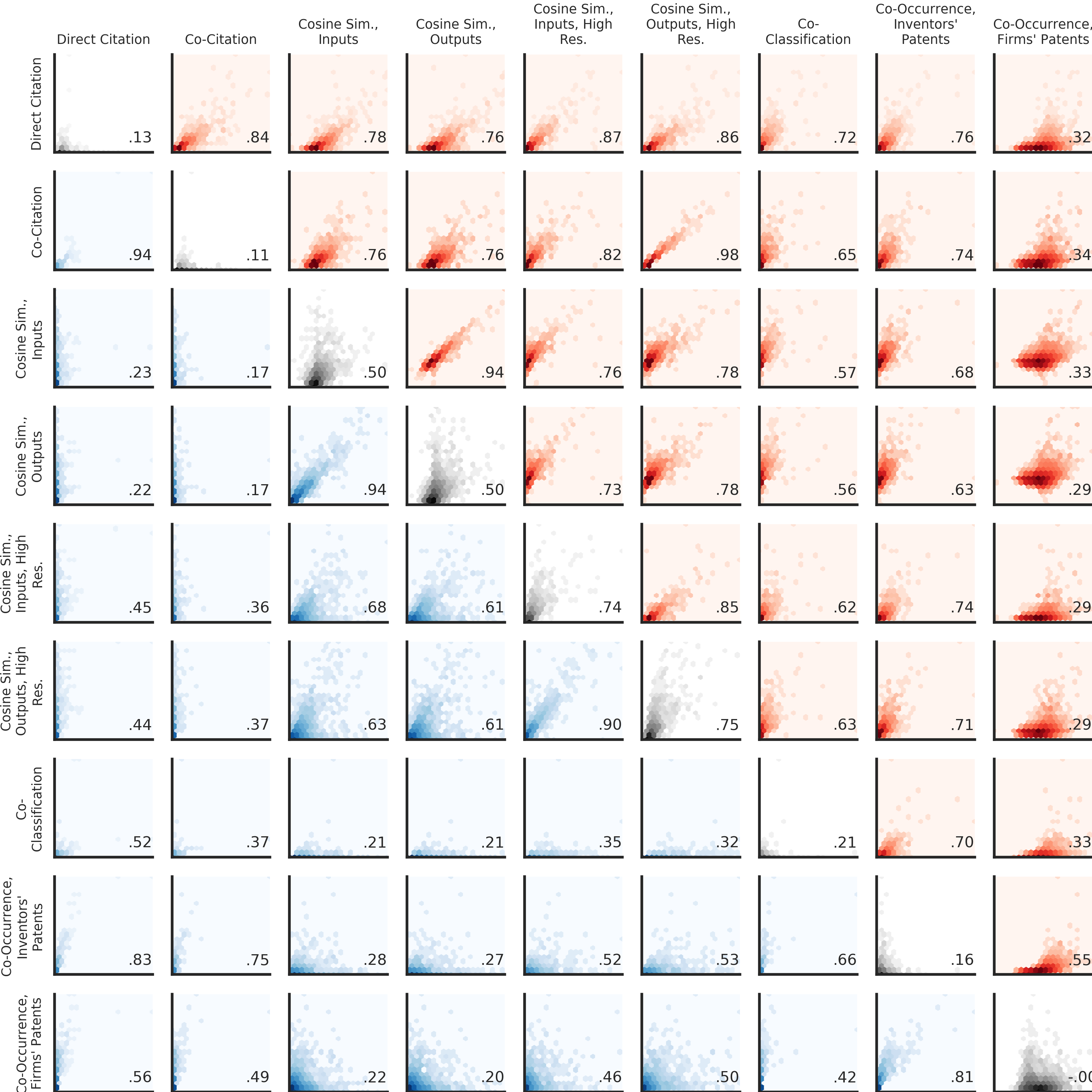} 
\caption{\textbf{Comparison of link weights across different proximity measures, as heat maps.} As Fig. \ref{Measure_Comparison_Scatter_Plots}, but expressed a heat map of point density. Blue: empirical networks. Red: normalized networks. Black: empirical vs. normalized link weight for each measure compared to itself (x-axis: empirical, y-axis: normalized). Text: Pearson correlation.
}\label{Measure_Comparison_Hex_Plots}
\end{figure*}

\section{Randomization of Co-Classification and Co-Occurrence Data}\label{Cooccurrence_Methods}
 The process for controlling for the number of occurrences of each class and the number of associations of each patent, inventor or firm is the same principle as used in the patent citation network: many randomized control versions of the empirical data are created.  In the randomized controls, the number of occurrences of each class and the number of associations of each patent, inventor or firm is preserved, but the assigning of technology classes to patents, inventors or firms is otherwise random. This goal is practically accomplished by expressing the co-classification or co-occurrence data as a bipartite graph, in which technology classes are one type of node and they form connections to another type of node, be that patents, inventors or firms. Randomized controls are then created by shuffling the bipartite network's links, but preserving each node's degree. The bipartite network is then projected into a single-mode network, the co-classification or co-occurrence network. The resulting one-mode network is then compared to the same information from the empirical, unrandomized data.

The bipartite network is  equivalent to a binary matrix, with patents, inventors or firms forming the rows and technology classes forming the columns. Reshuffling the bipartite network is equivalent to creating random versions of the binary matrix, with each row and column having the same sum. Bottazzi and Pirino described how creating random controls in this way for co-occurrence data can markedly alter the interpretation of empirical co-occurrences \cite{Bottazzi2010}. We extend the shuffling technique they used with a reshuffling method designed with bipartite networks in mind \cite{Gobbi2014}. These methods analytically determine the number of rewires necessary to make on the original bipartite network in order to effectively take an unbiased sample from the set of networks with the same degree sequence. We used the BiRewire software package to first calculate the necessary number of rewires for each of the bipartite networks we examined, and then to rapidly execute the rewires \cite{Gobbi2015}.

We created randomized controls that preserved temporal changes in class' popularity and inventors' or firms' activity by treating each year of data as a separate bipartite network. Each year was rewired independently, and then all years were combined to create the final randomized control to which the empirical network was compared.

Preserving temporal behavior introduces information which is not typically considered in the analysis of co-occurrence data. It also introduces additional complexity. Consider an inventor who was active in one technology class: \textit{Hats}. This inventor was active for 10 years, patenting each year, each time in the technology class for \textit{Hats}. If we identify this inventor as ``occurring" in \textit{Hats} each year, and randomize each of the 10 years individually, it is likely that across the 10 randomized years the inventor will then occur in  many different technology classes. When we combine the 10 randomized years back together, we would then observe that the inventor occurred in many technology classes, perhaps 10, which is far more than the 1 class in which the inventor was actually active. By marking the inventor as ``occurring" in every year individually, our randomization will thus break the  basic task of preserving the number of classes the inventor was active in. Therefore, any time an inventor or firm is awarded patents in a class in multiple years, we have a problem.

The solution to the problem of an inventor or firm ``occurring" over multiple years is to not consider the entity as occurring over multiple years. Instead, the inventor or firm is considered to occur in each class only once, in a single year. In this way, randomizing each year individually cannot increase the total number of classes that an entity associates with in its history. After randomization of individual years and combining them together, the number of classes per entity and entities per class will still be preserved. 

For the purpose of inventors or firms patenting in technology classes, the most salient year to mark the entity as occurring in a class is the \textit{first} year that entity patented in that class. This is particularly relevant for controlling for phenomena like popularity-chasing; if a firm only enters technology domains because they are popular, that does not provide more information about how related technology domains are. We thus mark each inventor or firm as occurring in a technology class when they first entered into that class, and compare to randomized controls that preserve the timing of the entries.

\subsection{Preserving Temporal Information Markedly Affected Firm Measures, but not Others}
\begin{figure}[htbp!]
\centering
\includegraphics[width=\columnwidth]{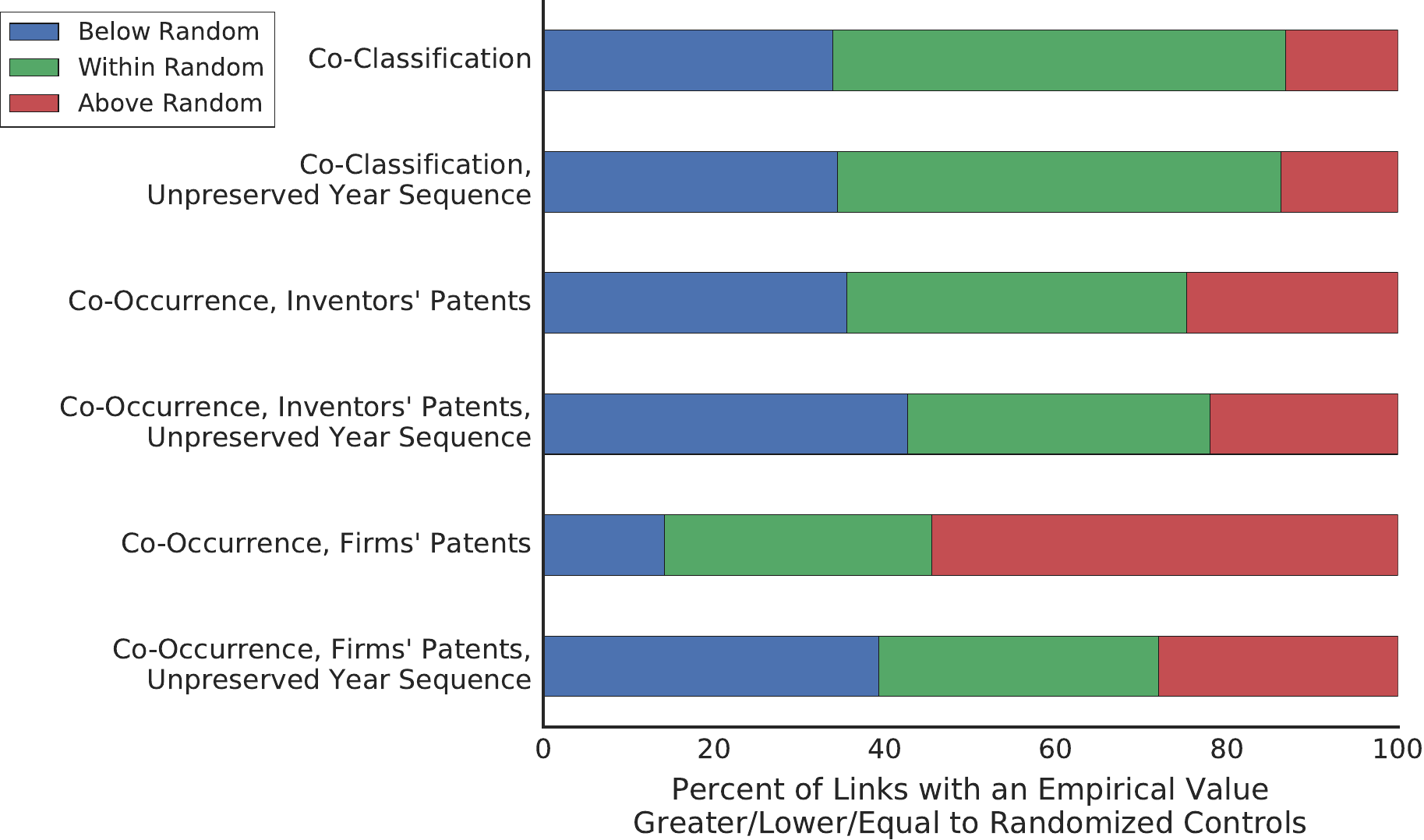} 
\caption{\textbf{Most proximity measures were only modestly affected by preserving yearly history.} The exception was in Co-Occurrence, Firm, which showed a marked change in the number of links that were stronger than chance.}\label{CoOccurrence_Year_Preservation_Comparison}
\end{figure}

Preserving the year sequence in the randomized controls had only a modest effect on the measured normalized network, for most measures (Fig. \ref{CoOccurrence_Year_Preservation_Comparison}). However, the Co-Occurrence, Firm measure was markedly altered by preserving the year sequences. Without preserving year sequences, the Co-Occurrence, Firm network had \textasciitilde{}25\% of its links stronger than random chance, closer to the fraction observed in the other proximity measures. By preserving the year sequences in the randomized controls and normalizing out such phenomena as popularity-chasing, the normalized Firm, Co-Occurrence network had \textasciitilde{}50\% of links stronger than chance.

By preserving temporal effects, the Co-Occurrence of technology classes in firms' patent portfolios were found to be generally more frequent than chance. Thus, by creating randomized controls that had more features in \textit{common} with the empirical data (the temporal sequence), the empirical data appeared more \textit{unusual}. This may seem counter intuitive, and so we provide some intuition here. Consider two statements:
\begin{quote}
1. ``I am a human, and I speak Mandarin Chinese". 
\end{quote}
This is unusual, but not that unusual. A randomly selected human has about a 1/7 chance of speaking Mandarin Chinese.

\begin{quote}
2. ``I am an Italian human, and I speak Mandarin Chinese".
\end{quote}
This is very unusual. A randomly selected Italian human has a much lower chance of speaking Mandarin Chinese.

Thus, by adding additional constraints to the randomized controls, we generate controls that are more like the empirical sample (human vs. Italian human), but the empirical sample is now more different from the controls.

In our case, we generate randomized controls that can either:
\begin{enumerate}
\item freely associate classes with firms, regardless of firms' histories
\item must associate classes with firms only when the firms entered a new class
\end{enumerate}

Consider a class, \textit{Hats}, that had some level of popularity $P$ across all of history, but during some periods of history had a much smaller popularity, $p$. Using method 1, randomized controls will match up a firm with \textit{Hats} at the rate $P$. However, it is possible that a specific firm only entered a new class at the moment in history when \textit{Hats} had the diminished popularity $p$.  Method 1 is blind to this fact. However, using method 2, randomized controls will match up the firm with \textit{Hats} at the rate $p$. The specific firm's entry into \textit{Hats}, then, appears more unusual using method 2 than method 1, because $p<P$. 

Therefore, using randomized controls that preserve the yearly sequence of firms' entries can identify temporal effects that make a firm's movement into a class appear more unusual. This method can then be scaled to a whole population of firms, to determine if their movements in aggregate are unusual. We can then measure whether the co-occurrence of two classes in firms' portfolios is unusual, i.e. different from that expected by chance.

\section{Co-Occurrence, Country Data}\label{cooccurrence_country}
\begin{figure}[htbp!]
\centering
\includegraphics[width=\columnwidth]{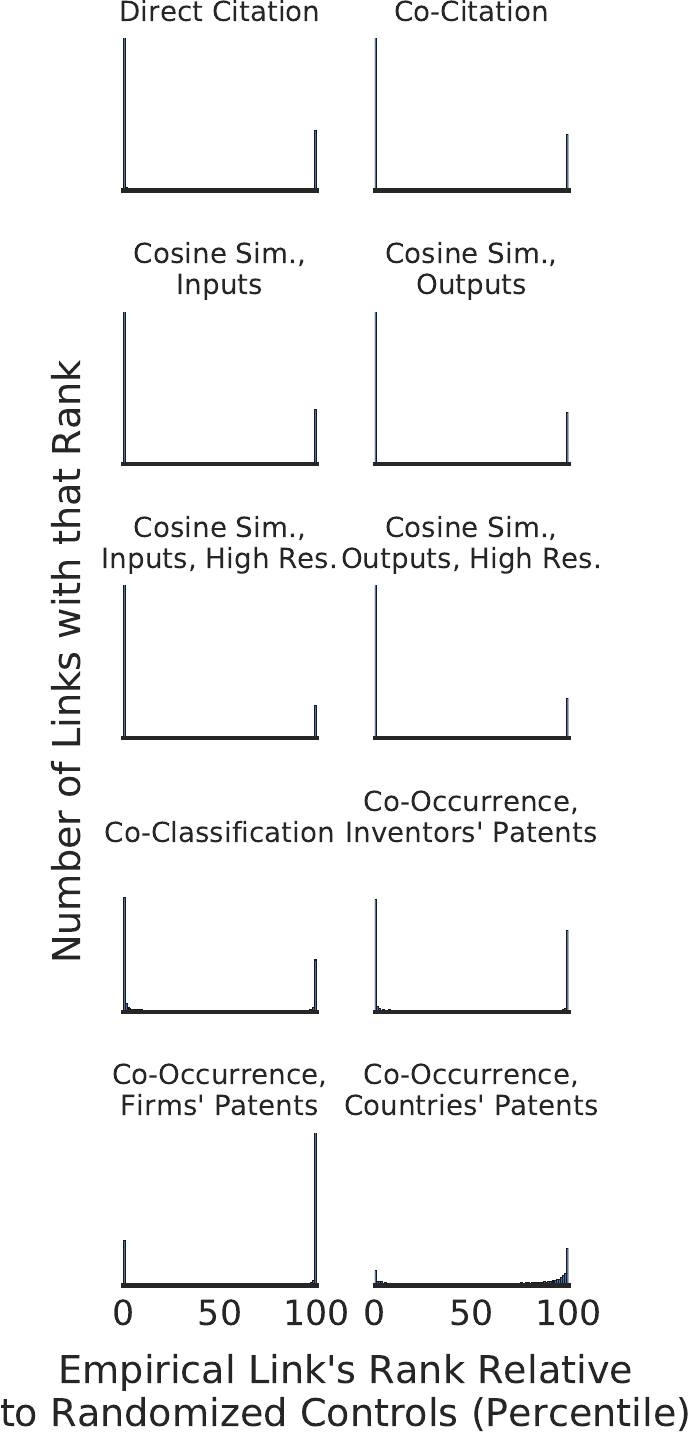} 
\caption{\textbf{All proximity measures found that most technology proximity links were very different from randomized controls, except Co-Occurrence, Country.}}\label{p_value_histograms}
\end{figure}

The empirical proximity links between technology classes had values typically much higher or lower than all 1,000 randomized controls, across all measures of proximity (Fig. \ref{related_unrelated_percentages}). We measure this phenomena in more detail by expressing each empirical link's value as a percentile rank, relative to the randomized controls. Fig. \ref{p_value_histograms} shows the histograms of the empirical links' ranks, for each proximity measure. For the nine proximity measures reported in the main text, the majority of links were lower than all randomized controls (rank 0) or higher (rank 100). For eight of the proximity measures, rank 0 links outnumbered rank 100 links. The exception was the Co-Occurrence, Firm network, in which rank 100 links were more common than rank 0 links.

We analyzed the patenting histories of countries to create a Co-Occurrence, Countries network, analogous to the networks created from Co-Occurrence, Inventor and Co-Occurrence, Firm measures. The country data was similar to the firm data, in that rank 100 links were more common than rank 0 data (Fig. \ref{p_value_histograms}, lower right panel). However, the vast majority of links between technology domains were between ranks 0 and 100, meaning they had values within the range expected by chance (covered by the 1,000 randomized controls). It is possible that with additional country data these links would prove to be significantly different from chance. However, with less than 200 countries, the co-occurrence data did not provide a sufficiently strong signal to assert that country's invention portfolios combined many technology classes at rates different from random chance.

\section{Analytic Approximations of Randomized Controls}\label{analytic_approximations}
\subsection{Expected Number of Citations}
The expected value of citations between any pair of classes and its standard deviation can be conveniently approximated analytically by exploiting the statistical properties of our randomization process. The process can be seen as a sum of random variables $X_{t,lag}$, one for each citing-cited year ${t,lag}$ pair that describe the possible relationship between a given citing and cited class. The citation swapping procedure can be described as sampling a number of citations $n_{citing_{t,lag}}$ without replacement out of a population $N$ of \textit{swappable} citations that respect the required constraints, in which there are exactly $K$ citations directed toward the given cited class. Therefore, for each citing-cited year pair ${t,lag}$, the expected number of citations between a citing and a cited class behaves like a hypergeometric random variable $X_{t,lag}$. As such, the total expected number of citations for a given pair of citing-cited classes is described by the sum ${C_{citing,cited}}$ of hypergeometric random variables ${X_{t,lag}}$ with different number of trials \textit{n}, population size $N_{t,lag}$ and number of successes $K_{cited_{t,lag}}$. It follows that the expected value $E(C_{citing,cited})$ is approximately equal to

\begin{multline}
E(C_{citing,cited}) \sim \sum_{\forall t \in T_{citing}}\sum_{lag=0}^{t-1976} \bigl[ n_{citing_{t,lag}} \\
* p(connection)_{cited_{t,lag}} \bigr]
\end{multline}\label{expected_value_citation}

Where $n_{citing_{t,lag}}$ is the number of citations made by the given citing class to be reshuffled for each random variable $X_{t,lag}$ (i.e. the number of trials). The probability of swapping any of these citations with anyone connecting to a patent belonging to the given cited class is

\begin{equation}
p(connection)_{cited_{t,lag}} = \frac{K_{cited_{t,lag}}}{N_{t,lag}}
\end{equation}\label{probability_connection}

More specifically, for each citing-cited year pair ${t,lag}$ the number of trials $n_{citing_{t,lag}}$ is equal to

\begin{equation}
n_{citing_{t,lag}} = C_{citing_{t}} p(lag)_{citing_{t,lag}} p(outward)_{citing_{t,lag}}
\end{equation}\label{number_of_trials}

Where $C_{citing_{t}}$ is the number of citations made by patents granted in year \textit{t} belonging to the given citing class, $p(lag)_{citing_{t,lag}}$ is the probability that they cite patents granted in the year ${t-lag}$ and $p(outward)_{citing_{t,lag}}$ is the probability that they cite patents belonging to a class different from the one of origin. The latter two are indexed by $citing$, $t$ and $lag$ because, as we have shown in the panels of Fig. \ref{Empirical_Normalized_Networks}, there is a large variability across classes and time. It follows that the standard deviation $\sigma_{citing,cited}$ of ${C_{citing,cited}}$ is approximately equal to

\begin{multline}\label{standard_deviation_citation}
\sigma_{citing,cited} \sim 
\sqrt{ 
\sum_{\forall t \in T_{citing}}\sum_{lag=0}^{t-1976} \bigl[ n_{citing_{t,lag}}} \\
\shoveright{* p(connection)_{cited_{t,lag}}} \\
\shoveright{* (1-p(connection)_{cited_{t,lag}})} \\
* \frac{N_{t,lag}-n_{citing_{t,lag}}}{N_{t,lag}-1} \bigr]  
\end{multline}

When $N_{t,lag}$ is large and $n_{t,lag}$ is small compared to it, then the fraction in equation \ref{standard_deviation_citation} approaches unity. Therefore, the $\sigma_{citing,cited}$ can be approximated by the standard deviation of a binomial distribution. This is particularly handy if one would like to have an analytic solution for the p-values of the empirical proximity. In fact, the distribution of the sum of hypergeometric random variables with varying number of trials and probability of success has no closed form solution. However, the sum of binomial random variables with different $n$ and $p$, can be seen as the the sum of Bernoulli random variables with different probabilities and is, therefore, described by the Poisson binomial distribution (a.k.a. Bernoulli-Poisson distribution) \cite{Hong2013, Fernandez2010, Chen1997}. Recently, it has been shown that the Poisson-binomial cumulative and probability distribution functions have exact closed-form solutions and accurate refined normal approximations \cite{Hong2013}. Based on the equations discussed here it is straightforward to derive the expected value of Co-Citation and Cosine Similarities between classes by using the joint probability distribution of citations from patents to classes and the cosine value of the vectors of expected received citations for any given class pairs.
 
\subsection{Nature and Quality of the Analytic Approximation}
\begin{figure*}[htbp]
\centering
\includegraphics[width=\textwidth]{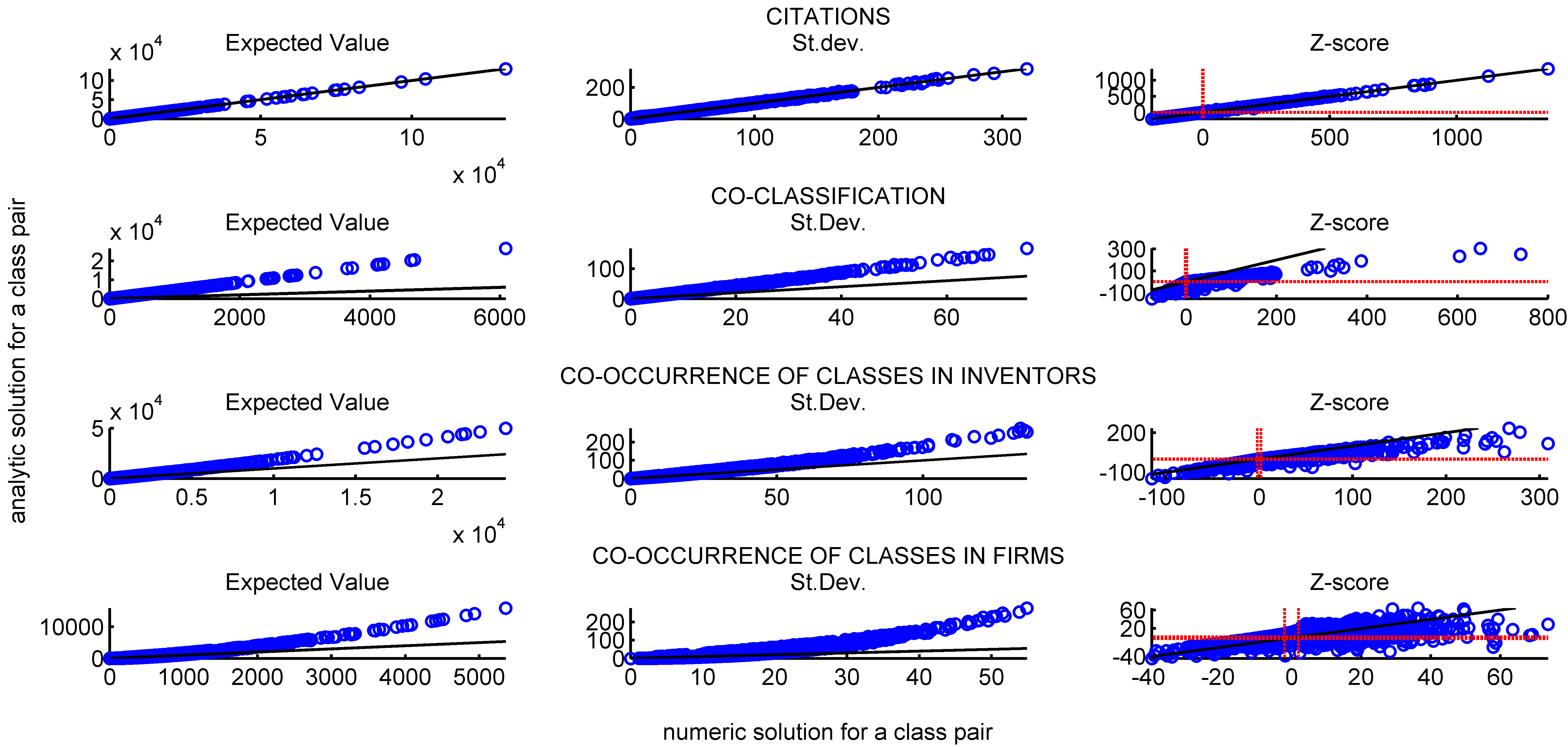} 
\caption{\textbf{Comparison of analytic and numeric solutions of the expected value (first column), standard deviation (second column) and z-scores (third column) of direct citations (first row), co-classification (second row), co-occurrences of classes in inventors' patenting histories (third row) and in firms histories (fourth row).} One data point per class pair. If the analytic and numeric solutions would agree, all data points would lay on the black solid lines. This only happens for direct citations. Red shaded lines in the z-scores panel highlights values of the z-scores equal to 2 and -2, i.e. a possible threshold of statistical significance for proximity and distance based on a normal approximation. Inference based on the analytic solutions would cause both type I and II errors for co-classification and co-occurrences. 
}\label{Comparison_ANAvsNUM}
\end{figure*}

The solutions for $E(C_{citing,cited})$ and $\sigma_{citing,cited}$ reported above are excellent approximations of numeric solutions  for the number of citations between classes, as provided by our randomization process (Fig. \ref{Comparison_ANAvsNUM}). The same approach could be applied to predict the numeric solution of the expected value and variance of co-occurrences of classes in patents, inventors' and firms' patenting histories. In this case $n$ would be the number of classes in which a patent, an inventor or a firm have been inventing, $K$ would be the number of patents, inventors or firms that have been patenting in a given class and $N$ would be the total number of occurrences (i.e. of links) in a bipartite network of patents*classes, inventors*classes or firms*classes. However, the approximation would perform very poorly in this case. 

The source of the analytic approximation deviating from the real behavior is due to the binary nature of citation networks and bipartite occurrence networks. When one works with weighted networks, numeric solutions provided by randomization algorithms that preserve row and column sums of the adjacency matrix, and analytic solutions based on hypergeometric distributions fully agree. In contrast, with binary networks, analytic solutions based on hypergeometric random variables may considerably differ from numeric solutions. The source of the problem lays in the possibility of double counting associations. Suppose that we are measuring the expected number of citations between patents. Suppose also that patent $A$ cites patents $B$ and $C$ and that patent $D$ cites patents $B$ and $E$, and that we want to swap citations between patents. If we use a permutation algorithm (which are a popular choice for randomizing weighted networks) to randomize citations, we might incur in double counting. In our example, if we permute cited patents we might end up in a configuration in which patent $A$ cites patent $B$ twice and patent $D$ cites patents $C$ and $E$. This would obviously break the binary nature of patent citations networks, and make the model deviate from reality. If our example would now describe a fictitious weighted networks we would not be facing double-counting problems as the random realization of the network would just strengthen the link between node $A$ and node $B$. Algorithms like BiRewire randomize networks while automatically preserving row and column sums of the empirical adjacency matrix, but also avoiding the possibility of double counting associations (this is accomplished by repeatedly swapping associations within sub-matrices of four cells in which associations are only found in one of the two diagonals). Analytic approaches based on hypergeometric distributions provide the exact solutions for permutation algorithms and are blind to the possibility of double-counting. Therefore, if we use them to predict expected values and standard deviation of associations between nodes in binary networks, we incur a mistake.

When we deal with a large patent citation network, in which the in-degree distribution is extremely skewed (i.e. most of the nodes have a very low $K$), most of the patents cite several patents (i.e. $n$ is relatively large), and there are many links (i.e. $N$ is very large), the probability of selecting two citations to swap that will cause double-counting is very low. To understand this, suppose now that patent $A$ cites patents $B$ and $C$ (and therefore has $n=2$), that patent $B$ has been cited $K$ times and that there are $N$ citations in the network. We would face double counting of citations from patent $A$ to $B$ only if, during the randomization process, we would randomly pick the citations from $A$ to $C$ and swap it with another one directed to $B$. This will happen with probability $(1/n)*(K-1)/(N-n)$, which is very small. For these reasons, the hypergeometric-based analytic solution, designed to predict permutation algorithms, works very well in our case, even if we randomize our binary citation network by using a 2-by-2 sub-matrix diagonal swapping algorithm (Fig. \ref{Comparison_ANAvsNUM}), top row). However, for co-occurrence data, the situation is very different. 

The occurrence networks of patents-classes, inventors-classes and firms-classes have much fewer links than a patent citation network (i.e. $N$ is much smaller) and the occurrence of some classes is much more common that the appearance of citations to a given patent (i.e. $K$ is much larger compared to $N$ in our occurrence networks). Therefore, the probability of incurring in double counting, if we would use permutation algorithms, is much larger. Accordingly, 2-by-2 sub-matrix diagonal swapping algorithms, like BiRewire, must be used in this case and the misuse of hypergeometric-based analytic solutions to predict their outcome actually causes type I and II errors in the inference based on z-scores (Fig. \ref{Comparison_ANAvsNUM}, right column). For this reason, precise analytic solutions of the expected value and variance of occurrences of classes in patents, inventors' and firms' histories do not exist. We must therefore solely rely on the numeric solutions provided by our randomization method, to calculate reliable z-scores of technology proximity.        

\clearpage

\clearpage

\begin{figure}[]
\centering
\includegraphics[width=\columnwidth]{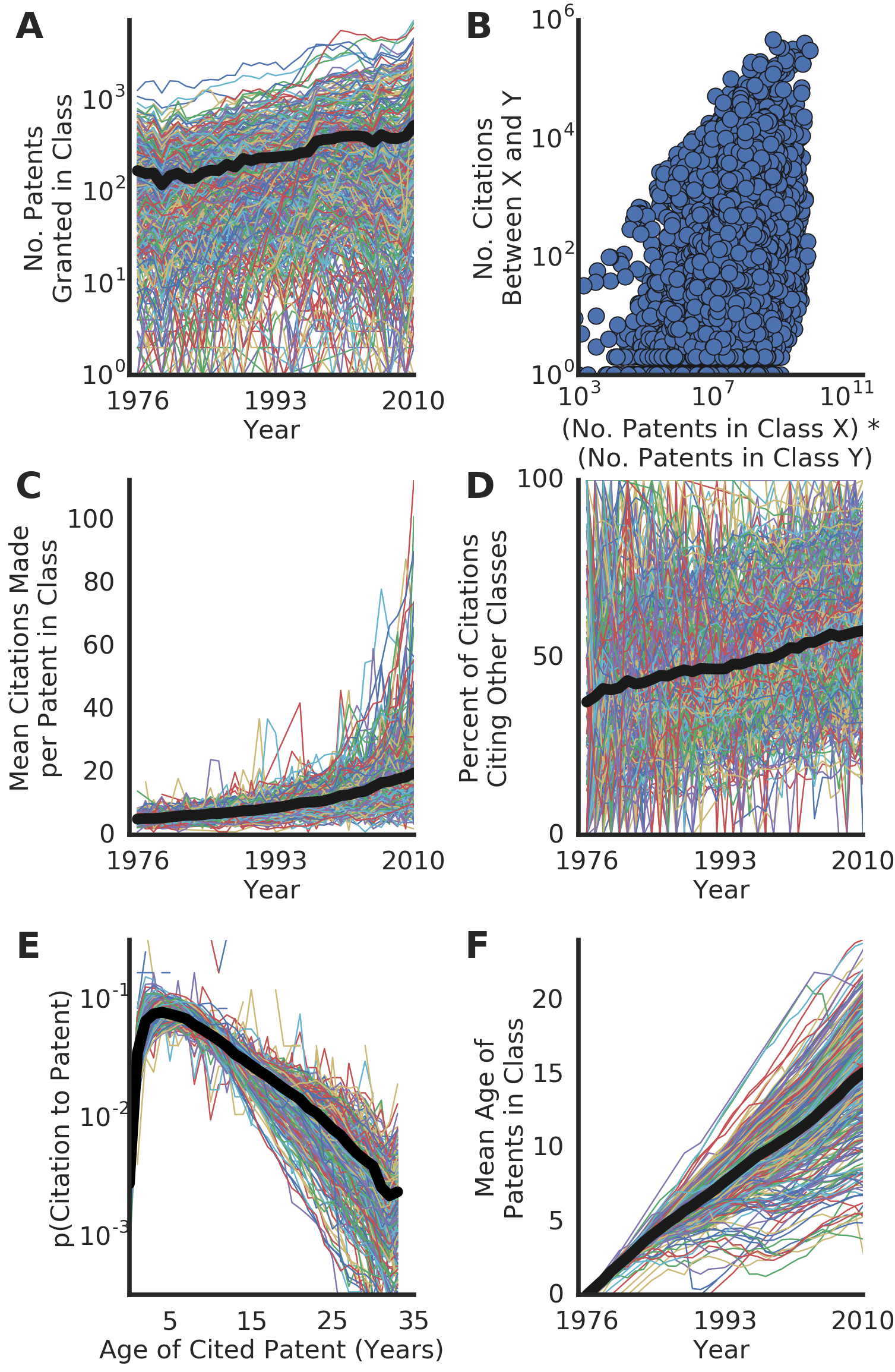} 
\caption{\textbf{The impinging factors affecting proximity measures, calculated using the USPC classification system.}}
\end{figure}

\begin{figure}[]
\centering
\includegraphics[width=\columnwidth]{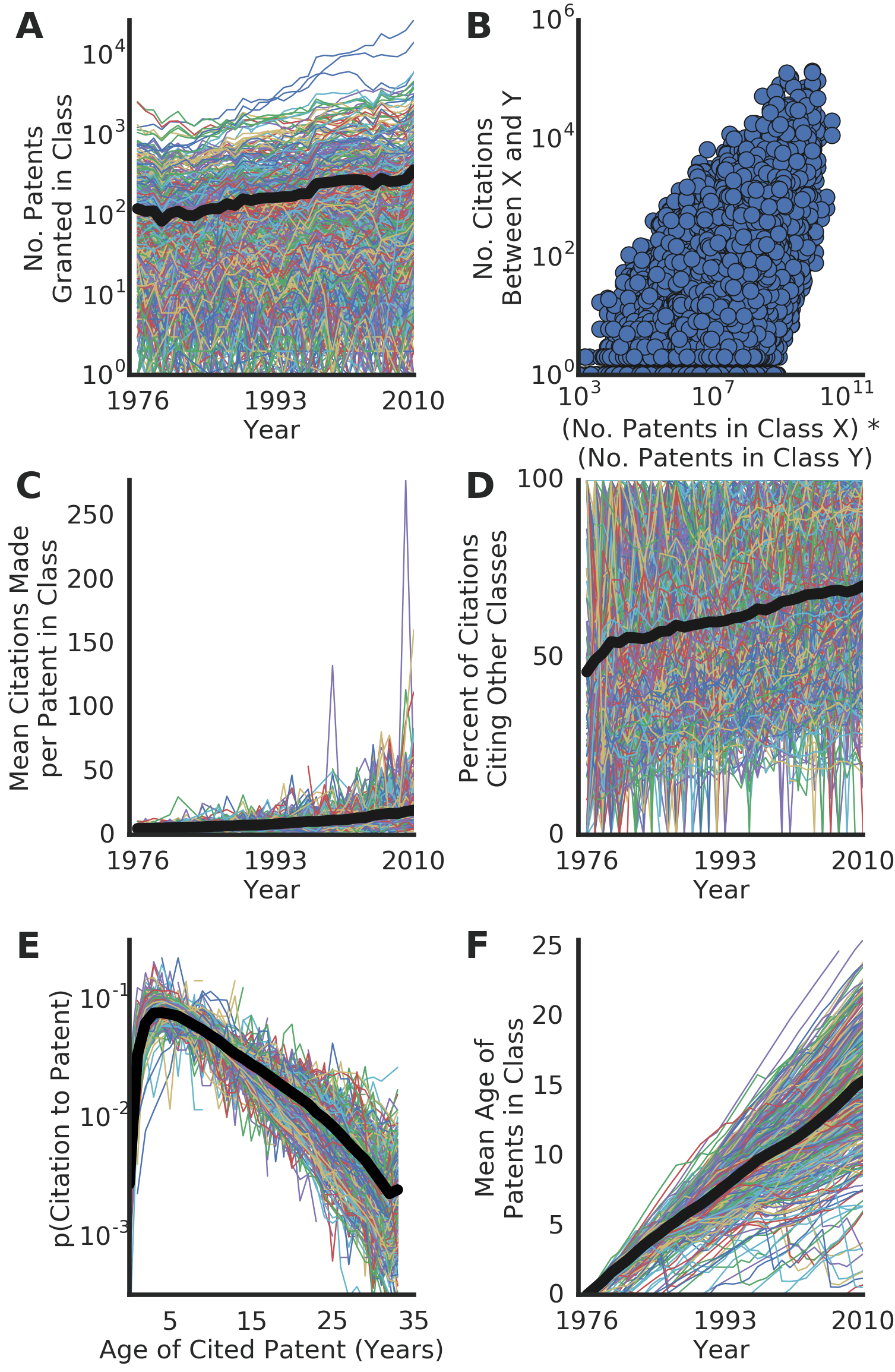} 
\caption{\textbf{The impinging factors affecting proximity measures, calculated using the IPC4 classification system.}}
\end{figure}

\clearpage

\begin{figure}[p!]
\centering
\includegraphics[width=\columnwidth]{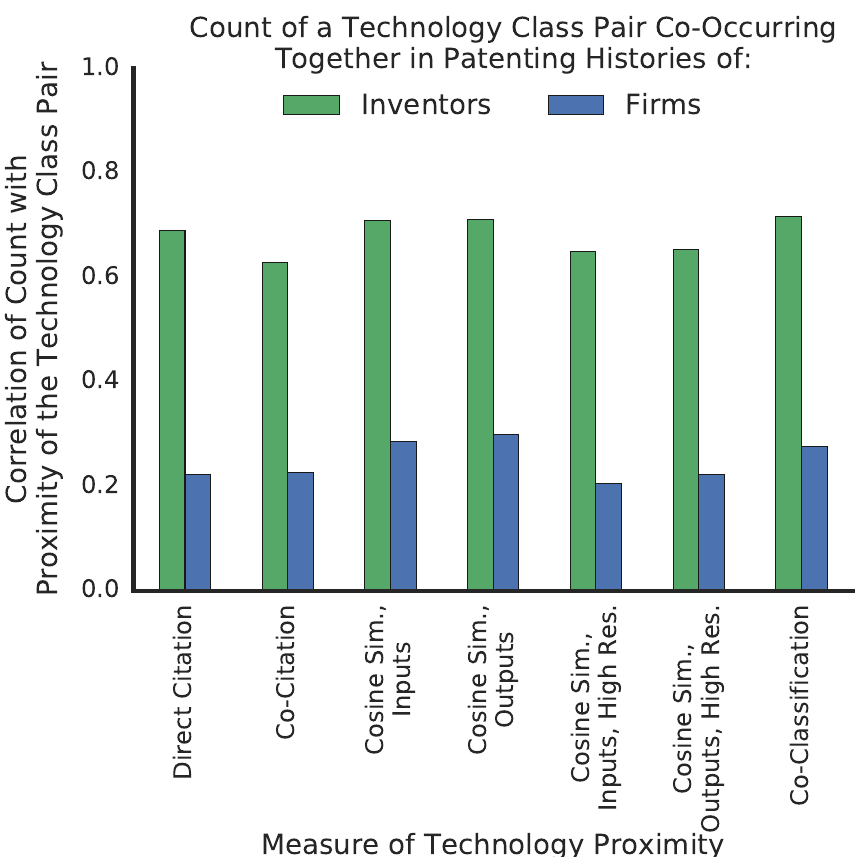} 
\caption{\textbf{As calculated using the USPC classification system, the normalized measures of technology proximity strongly correlated with the behavior of inventors, and modestly with the behavior of firms.}}
\end{figure}

\begin{figure}[p!]
\centering
\includegraphics[width=\columnwidth]{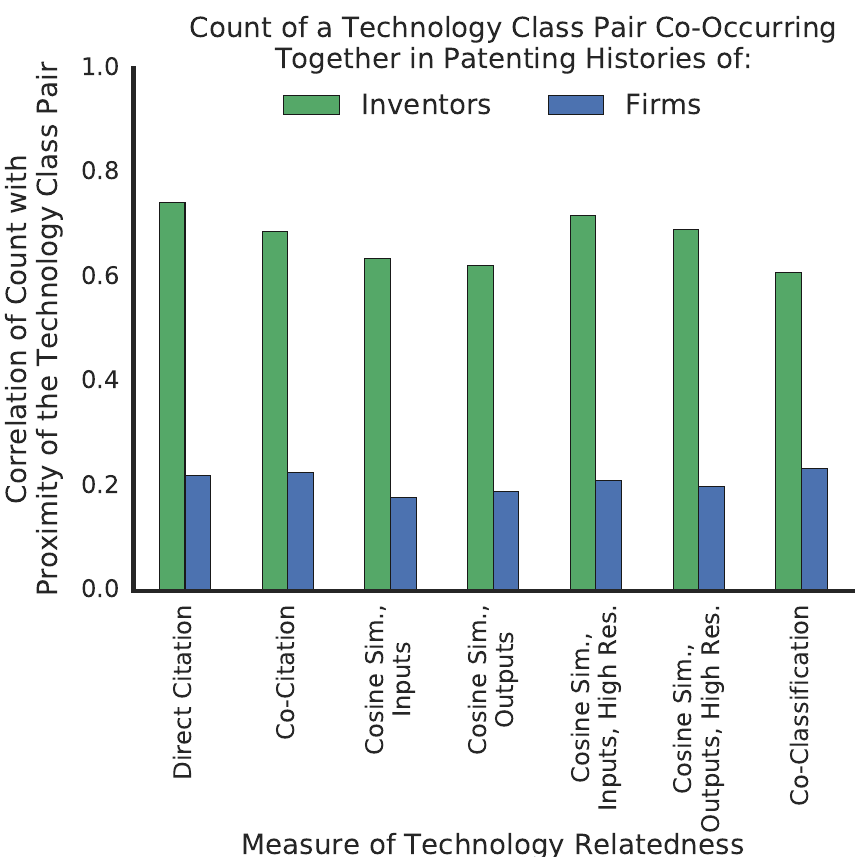} 
\caption{\textbf{As calculated using the IPC4 classification system, the normalized measures of technology proximity strongly correlated with the behavior of inventors, and modestly with the behavior of firms.}}
\end{figure}

\clearpage
\printbibliography

\newpage
\beginsupplement
\pagebreak
\clearpage

\begin{center}
\textbf{\large Supporting Information}
\end{center}
\section{Z-Score Inflation and Deflation of Different Proximity Measures}
All figures as Fig. \ref{z-score_deflation}. Note that cosine-based measures have different dynamics with how randomized controls' means and standard deviations alter with the number of patents in a pair of classes. However, since the mean and the standard deviation still change at different rates, their ratio still changes, which still leads to the change in the space of possible z-scores. Thus, correcting for the change in the space of possible z-scores is still necessary.

\begin{figure*}[p]
\begin{center}
\includegraphics[width=\textwidth]{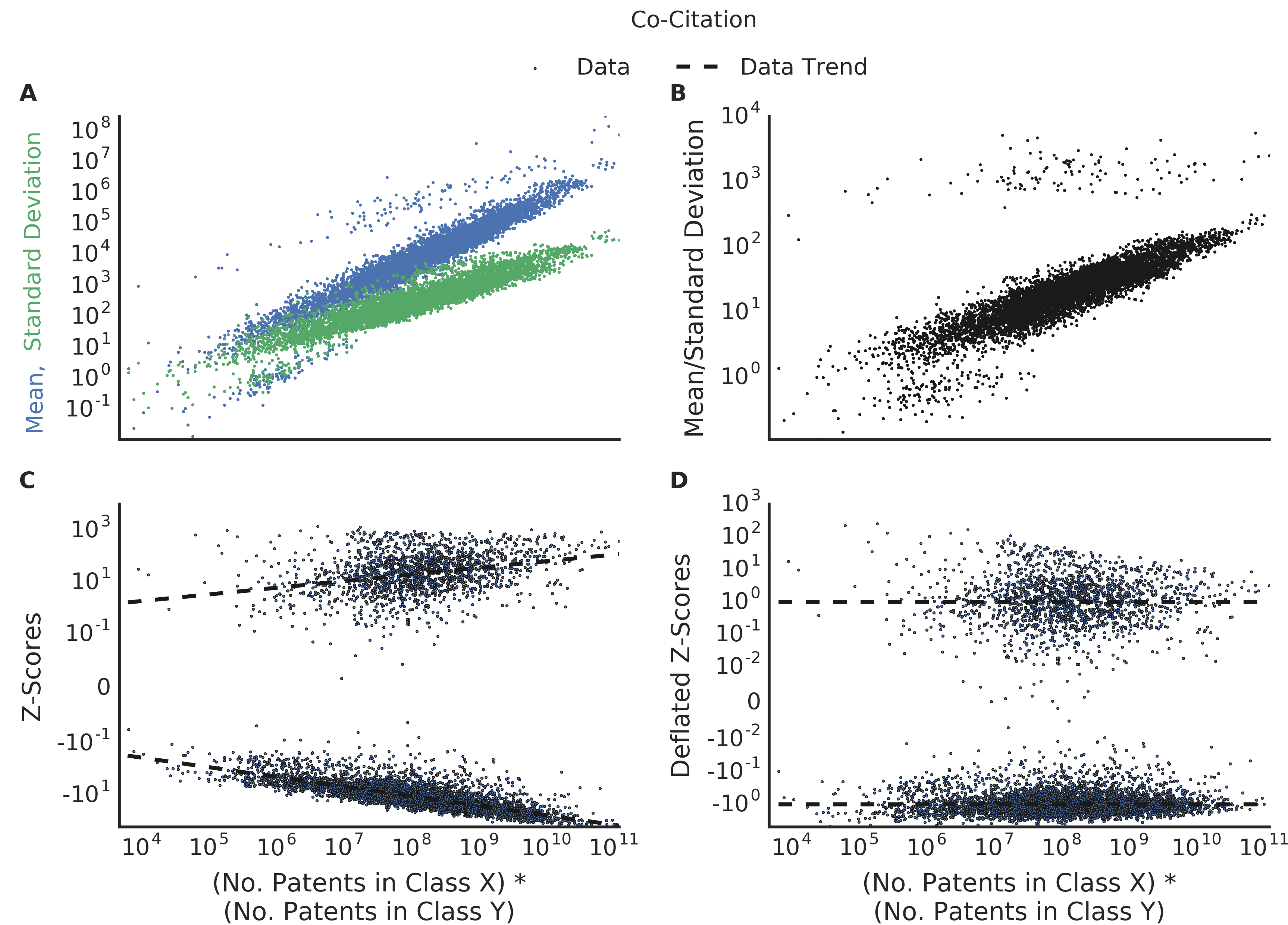} 
\end{center}
\caption{}
\end{figure*}

\begin{figure*}[]
\begin{center}
\includegraphics[width=\textwidth]{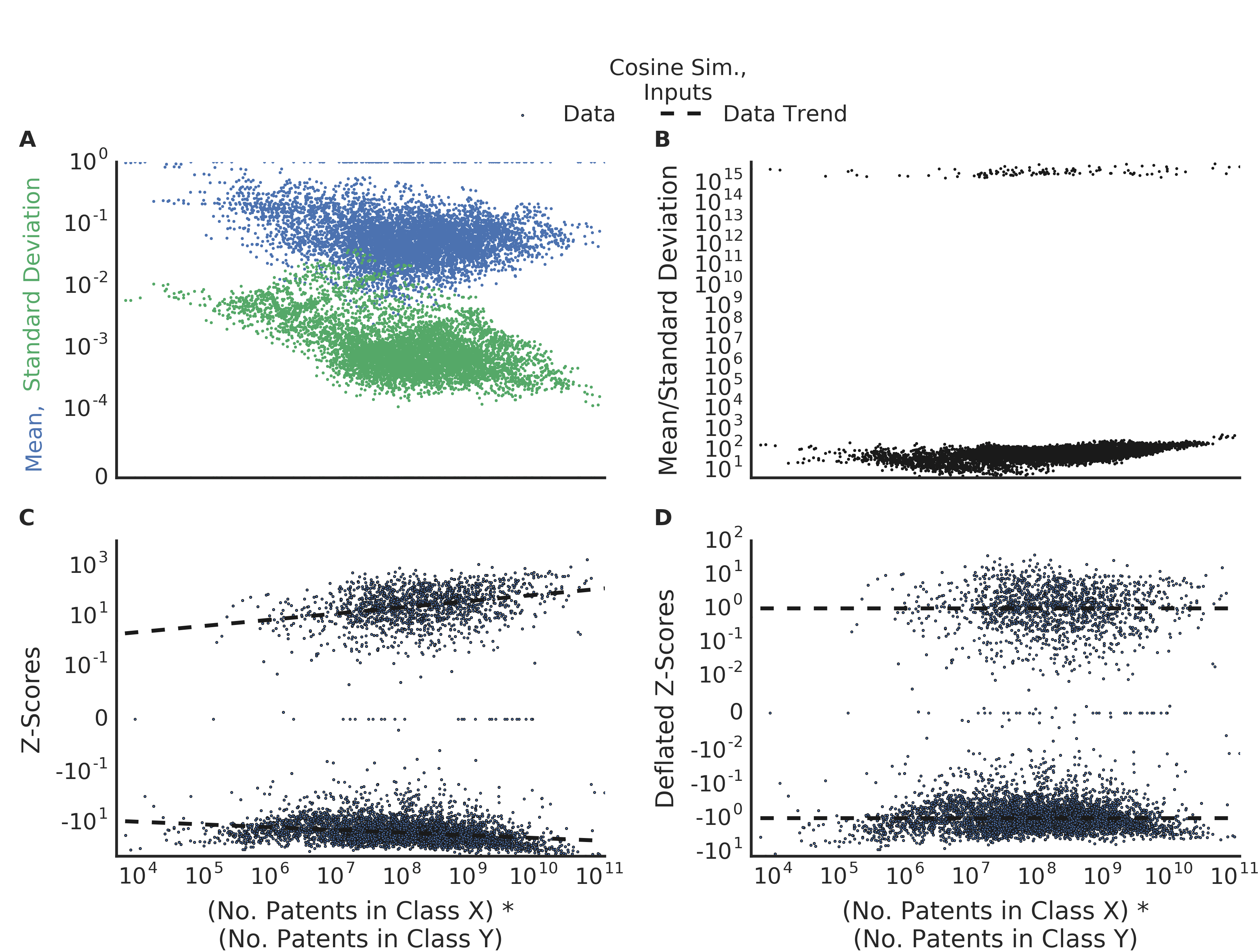} 
\end{center}
\caption{}
\end{figure*}

\begin{figure*}[]
\begin{center}
\includegraphics[width=\textwidth]{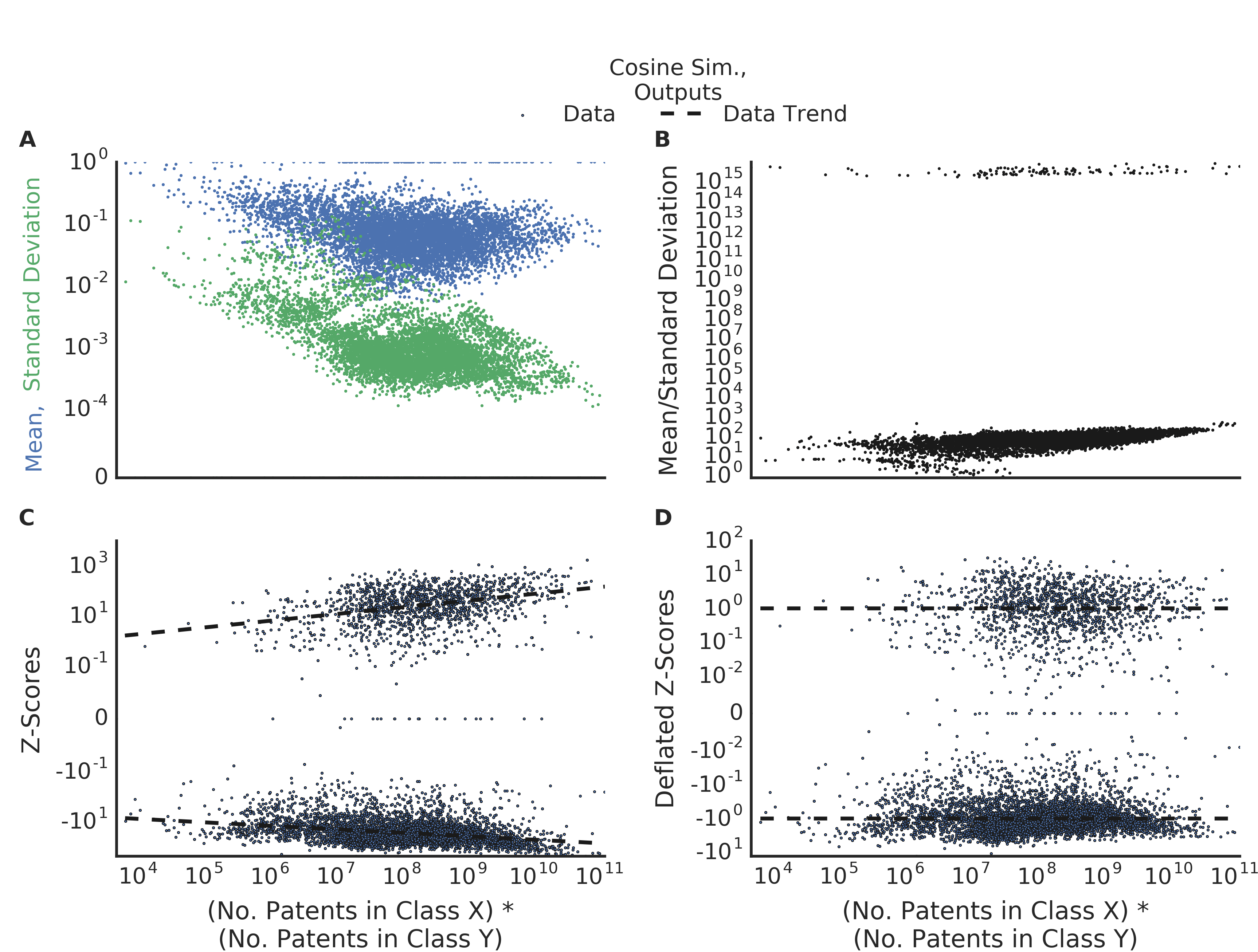} 
\end{center}
\caption{}
\end{figure*}

\begin{figure*}[]
\begin{center}
\includegraphics[width=\textwidth]{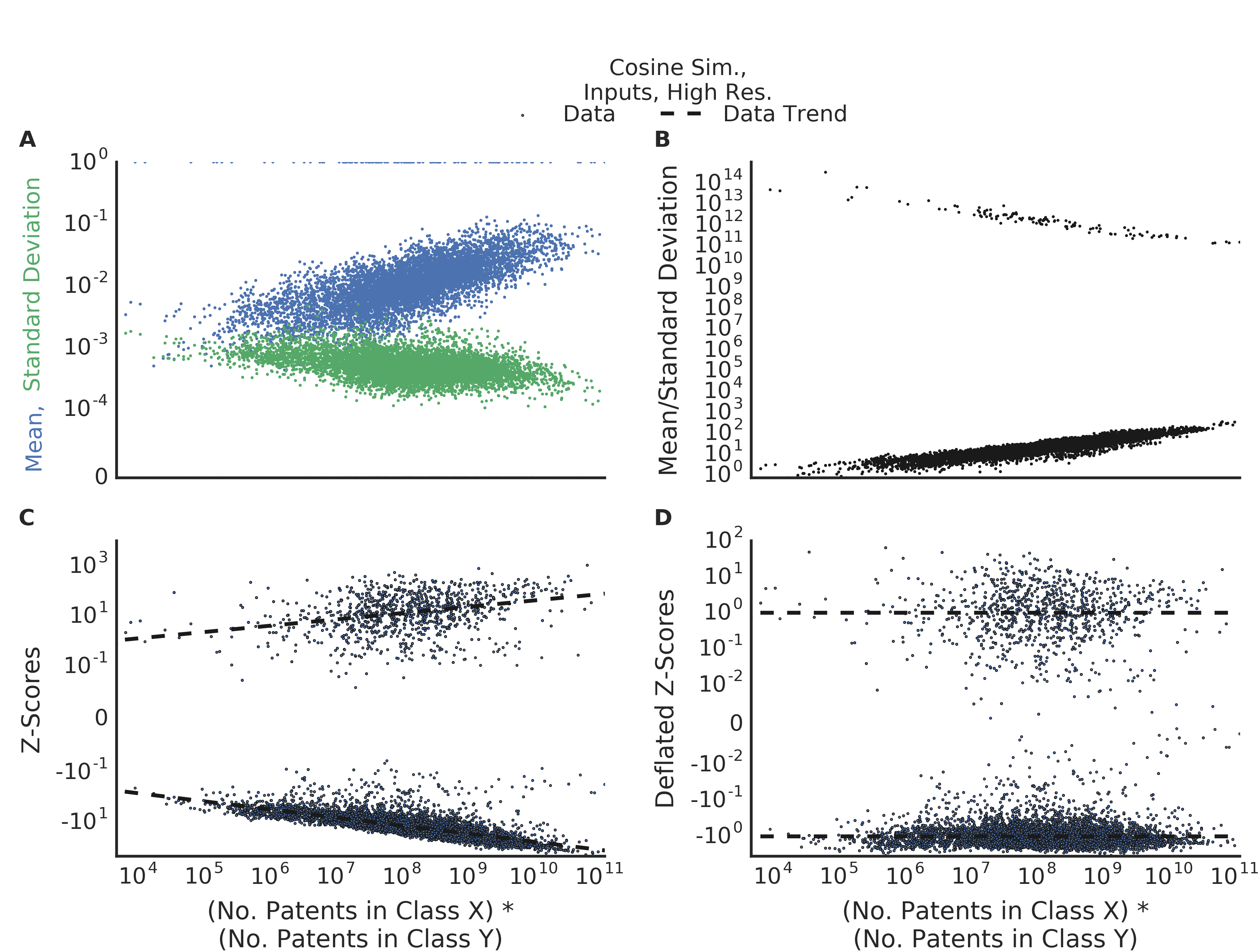} 
\end{center}
\caption{}
\end{figure*}

\begin{figure*}[]
\begin{center}
\includegraphics[width=\textwidth]{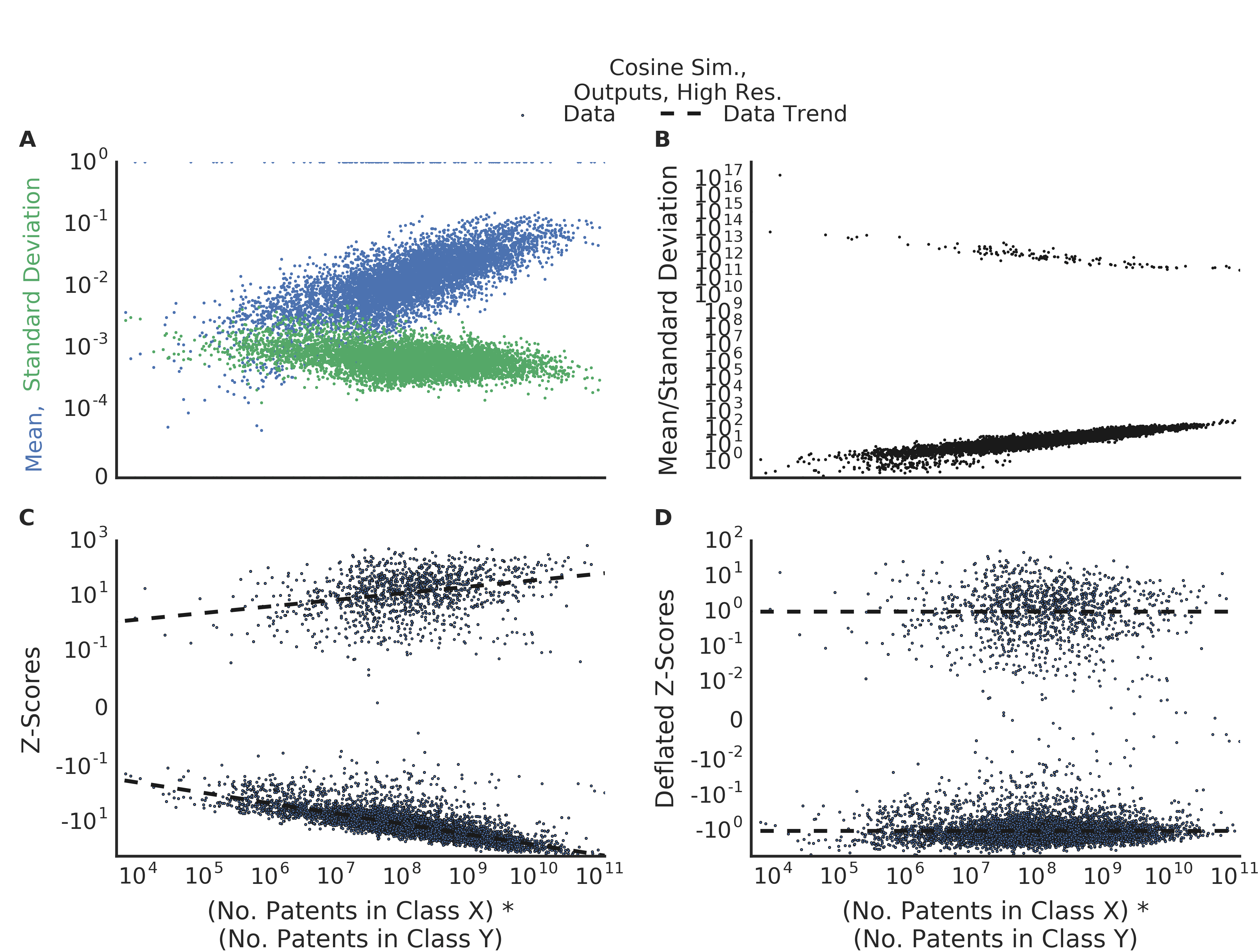} 
\end{center}
\caption{}
\end{figure*}

\begin{figure*}[]
\begin{center}
\includegraphics[width=\textwidth]{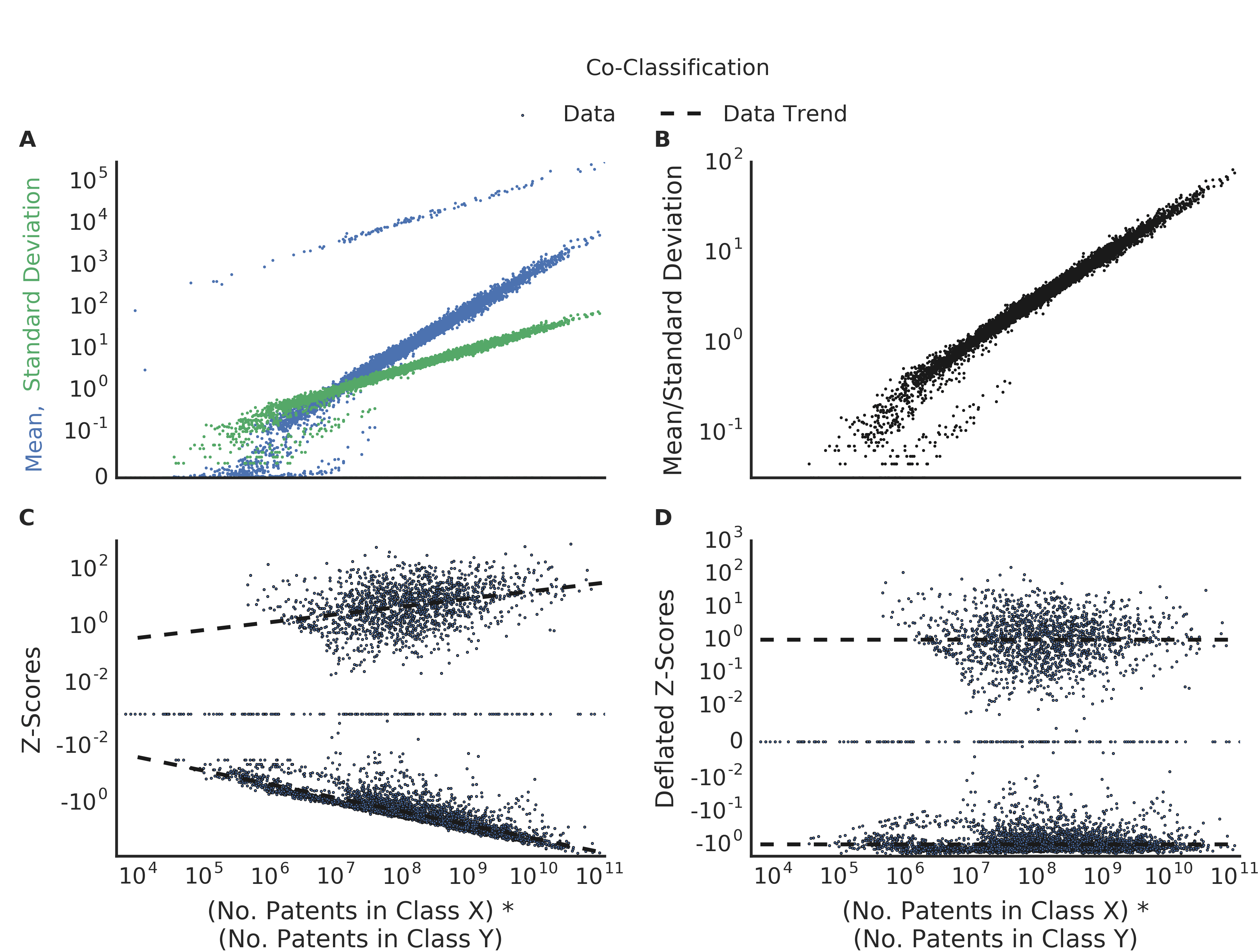} 
\end{center}
\caption{}
\end{figure*}

\begin{figure*}[]
\begin{center}
\includegraphics[width=\textwidth]{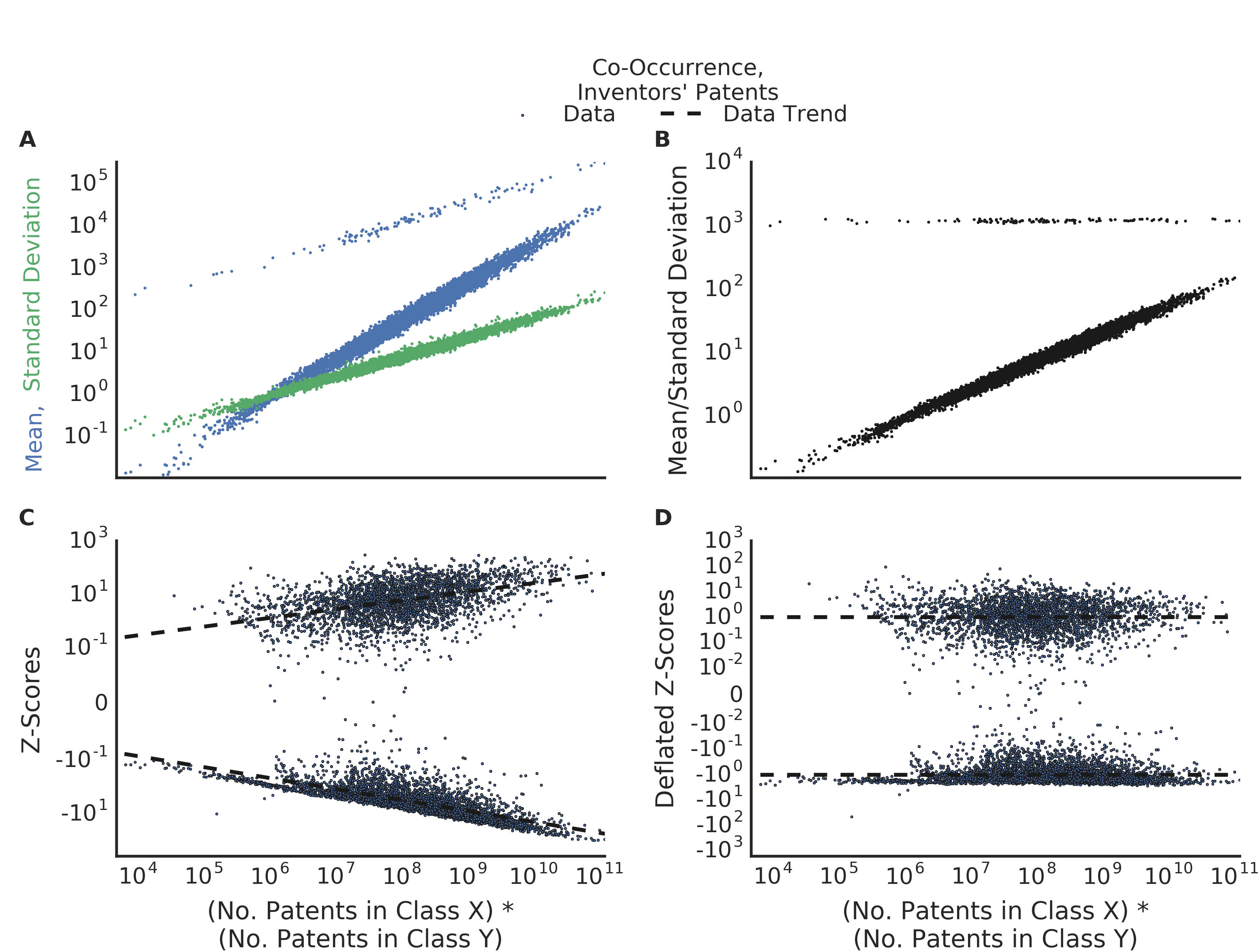} 
\end{center}
\caption{}
\end{figure*}

\begin{figure*}[]
\begin{center}
\includegraphics[width=\textwidth]{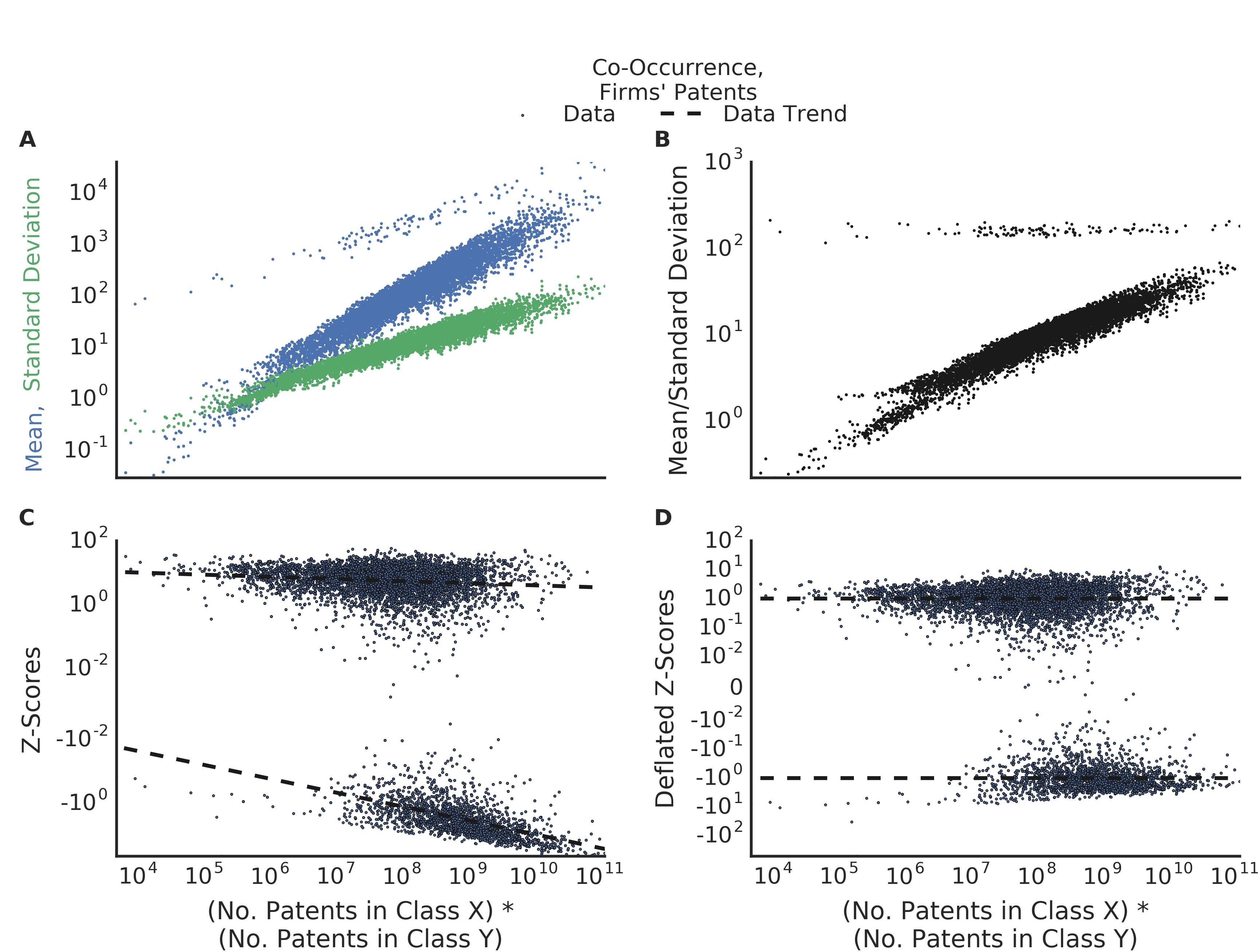} 
\end{center}
\caption{}
\end{figure*}

\end{document}